\RequirePackage{fix-cm}
\documentclass[twocolumn,epjc3]{svjour3}
\smartqed  
\RequirePackage{graphicx}
\RequirePackage[numbers,sort&compress]{natbib}
\RequirePackage[colorlinks,citecolor=blue,urlcolor=blue,linkcolor=blue]{hyperref}
\RequirePackage{amsmath, cases, empheq} 
\RequirePackage{amssymb}
\RequirePackage{amsfonts}
\RequirePackage{bm}
\RequirePackage{lineno}
\RequirePackage{comment}
\journalname{Eur. Phys. J. C}
\begin{document}

\title{Neutron source-based event reconstruction algorithm in large liquid scintillator detectors
}


\author{Akira Takenaka\thanksref{e1,addr1}
        \and
        Zhangming Chen \thanksref{addr2}
        \and
        Arran Freegard \thanksref{addr2}
        \and
        Junting Huang \thanksref{addr2}
        \and
        Jiaqi Hui \thanksref{addr1}
        \and
        Haojing Lai \thanksref{addr2}
        \and
        Rui Li \thanksref{addr2}
        \and
        Yilin Liao \thanksref{addr2}
        \and
        Jianglai Liu \thanksref{addr1, addr2}
        \and
        Yue Meng \thanksref{addr2}
        \and 
        Iwan Morton-Blake \thanksref{addr1}
        \and
        Ziqian Xiang \thanksref{addr2}
        \and
        Ping Zhang \thanksref{addr2}
}

\thankstext{e1}{e-mail: akira.takenaka@sjtu.edu.cn}


\institute{Tsung-Dao Lee Institute, Shanghai Jiao Tong University, Shanghai, China \label{addr1}
           \and
School of Physics and Astronomy, Shanghai Jiao Tong University, Shanghai, China  \label{addr2}
}

\date{Received: date / Accepted: date}

\maketitle

\begin{abstract}
We developed an event reconstruction algorithm, applicable to large liquid scintillator detectors, built primarily upon neutron calibration data.
We employ a likelihood method using photon detection time and charge information from individual photomultiplier tubes.
Detector response tables in the likelihood function were derived from americium-carbon neutron source events, 2.2~MeV $\gamma$-ray events from cosmic-ray muon spallation neutrons, and laser calibration events.
This algorithm can reconstruct the event position, energy, and also has the capability to differentiate particle types for events within the energy range of reactor neutrinos.
Using the detector simulation of the Jiangmen Underground Neutrino Observatory (JUNO) experiment as a large liquid scintillator detector example, we demonstrate that the presented reconstruction algorithm has a reconstructed position accuracy within $\pm$4~cm, and a reconstructed energy non-uniformity under 0.5\% throughout the central detector volume.
The vertex resolution for positron events at 1~MeV is estimated to be around 9~cm, and the energy resolution is confirmed to be comparable to that in the JUNO official publication.
Furthermore, the algorithm can eliminate 80\% (45\%) of $\alpha$-particle (fast-neutron) events while maintaining a positron event selection efficiency of approximately 99\%.
\end{abstract}

\section{Introduction}
\label{sec:introduction}
Large liquid scintillator detectors have played a crucial role in particle physics, particularly in neutrino physics. 
Recent examples include KamLAND~\cite{KamLAND:2002uet}, Daya-Bay~\cite{DayaBay:2012fng}, Double Chooz~\cite{DoubleChooz:2011ymz}, and RENO~\cite{RENO:2012mkc}, that made precise measurements of reactor neutrino oscillations, as well as Borexino~\cite{BOREXINO:2014pcl,BOREXINO:2018ohr,BOREXINO:2020aww}, which made comprehensive solar neutrino measurements.
Currently, several new-generation liquid scintillator-based neutrino detectors have begun or will begin taking data soon, such as Kamland-Zen~\cite{KamLAND-Zen:2022tow}, SNO+~\cite{SNO:2021xpa}, Jinping Neutrino Experiment~\cite{Jinping:2016iiq}, and JUNO~\cite{JUNO:2021vlw}.
These detectors use photomultiplier tubes (PMTs), mounted on the detector support structure, to detect the scintillation and Cherenkov photons produced in the liquid scintillator region due to the energetic, charged particles generated in neutrino interactions.
In this paper, we discuss a data-driven algorithm capable of reconstructing the event position, energy, and particle type of charged particles based on the photon detection times and the amount of detected photons at each photomultiplier tube.
As an application example of the algorithm, we developed and evaluated the performance of the algorithm using the detector simulation of the JUNO experiment~\cite{Lin:2022htc}; 
however, the method itself can be directly applied to events observed in the actual detector. \par
\begin{figure}
\includegraphics[width=0.48\textwidth]{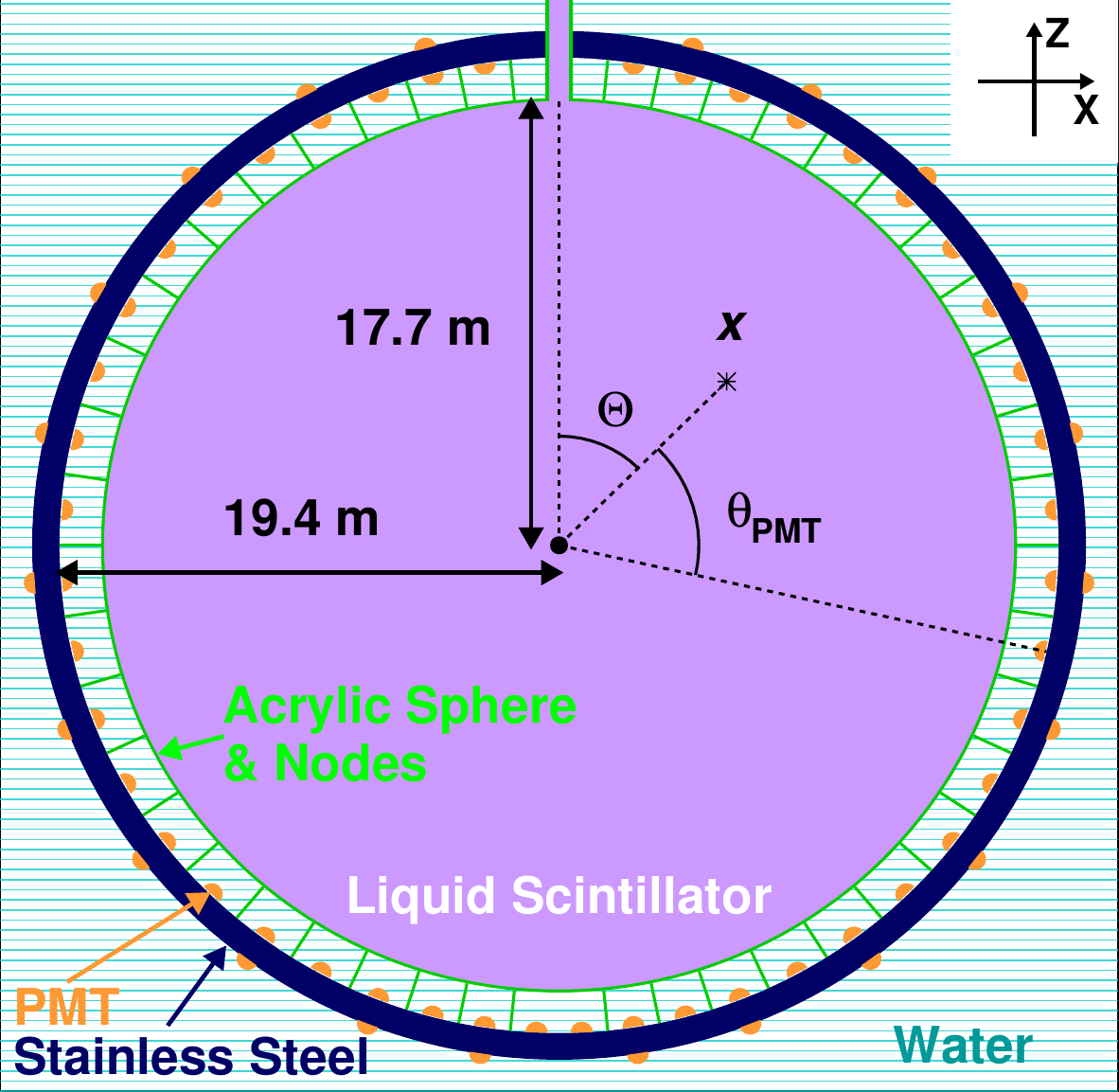}
\caption{
A schematic view of the JUNO detector, a 20-kton liquid scintillator volume, contained in a spherical acrylic vessel and viewed by PMTs mounted on the stainless steel support structure. 
The acrylic vessel and stainless steel structure are connected by 590 acrylic nodes (connecting bars), and the volume outside of the acrylic vessel is filled with purified water.
The detector $X$- and $Z$-axes are defined in the top right corner of this figure, and the origin of the coordinates is the center of the spherical liquid scintillator volume. 
The asterisk mark (*) in the figure represents the event vertex position, $\bm{x}=(R,\Theta,\Phi)$. 
The relative angle between the vertex position and PMT is expressed as $\theta_{\rm PMT}$. 
The radius of the spherical liquid scintillator volume is 17.7~m, and the distance between the detector center and PMT surface plane is approximately 19.4~m.
PMTs mounted on the outer surface of the stainless steel support structure help to detect cosmic-ray muons from the outside of the detector.
These PMTs are not involved in the reconstruction algorithm presented in this study.
}
\label{fig:detector}
\end{figure}

JUNO is a 20-kiloton liquid scintillator detector, currently under construction in Jiangmen, China. 
Its primary physics goal is to determine the neutrino mass ordering by precisely measuring the reactor neutrino energy spectrum from the reactors located roughly 52.5~km (on average) from the experimental site.
The detector will be equipped with 17,612 20-inch large PMTs (LPMT, 75\% photo coverage) and 25,600 3-inch small PMTs (SPMT, 3\% photo coverage) oriented towards the inner detector volume, which consists of a 17.7~m radius spherical acrylic vessel containing the 20 ktons of liquid scintillator, serving as the neutrino interaction target, as shown in Figure~\ref{fig:detector}.
Electron antineutrinos from the nuclear reactors interact with protons in the liquid scintillator volume via the inverse beta decay process;
$\overline{\nu}_{\rm e} {\rm p} \rightarrow {\rm e}^{+} {\rm n}$.
The positron deposits its kinetic energy in the liquid scintillator and annihilates with an electron, producing two 511~keV $\gamma$-rays at the annihilation point.
In addition, the emitted neutron thermalizes in the liquid scintillator and is eventually captured with a typical capture time of approximately 200~$\mu$sec.
In the liquid scintillator, neutrons are mostly (about 99\% of the time) captured by hydrogen nuclei, and a 2.2~MeV $\gamma$-ray is emitted from the de-excitation process of the resulting deuteron. 
In the event of a neutron being captured on carbon ($\sim$1\% of the time), one or two $\gamma$-rays are emitted with a total energy of 4.9~MeV.
To determine the neutrino mass ordering in JUNO, key points closely related to the event reconstruction algorithm are as follows:
\begin{enumerate}
\item An optimized energy resolution for these outgoing particles from the neutrino interaction, through correcting the event position-dependent non-uniform response of the detector.
Detector calibration and simulation studies~\cite{JUNO:2020xtj} indicated that the energy scale inside the detector could exhibit more than 10\% non-uniformity. 
This non-uniformity arises from variations in effective PMT coverage as a function of event position, as well as complex light propagation processes, such as light absorption, scattering within the detector medium, and reflection/refraction at the detector boundaries. 
Such non-uniform energy responses introduce additional spread in energy estimation, requiring the event reconstruction algorithm to remove these effects to achieve optimal energy resolution.
\item Detector fiducialization. 
One of the primary background sources in the reactor neutrino analysis in JUNO is accidental background caused by radioactive impurities contaminating the detector components. 
Past radioactive assessments of each detector component~\cite{JUNO:2021kxb} suggested that higher background event rates are expected near the boundary of the liquid scintillator, induced by detector components outside the liquid scintillator region. 
Additionally, particles produced near the boundary of the liquid scintillator are more likely to escape, leading to energy leakage. 
To eliminate backgrounds from external radioactive impurities and ensure accurate energy measurement, it is crucial to precisely reconstruct the event position and fiducialize the detector volume for physics analysis.
\item Background removal with pulse-shape discrimination.
Besides backgrounds from the ambient detector materials, $\alpha$-particles produced by radioactive impurities in the liquid scintillator and fast-neutrons generated by cosmic-ray muons also pose challenges in the reactor neutrino analysis. 
Pulse-shape information can aid in distinguishing these backgrounds.
\end{enumerate}
\par

The algorithm presented in this paper employs a maximum likelihood method, in which the position and energy of the particle are reconstructed by maximizing a defined likelihood function based on the charge and timing information observed by the PMTs (see Figure~\ref{fig:detector}).
The principle of this method has been widely employed in neutrino experiments, such as MiniBooNE~\cite{Patterson:2009ki}, Daya-Bay~\cite{Wen:2011zzb}, Super-Kamiokande~\cite{Super-Kamiokande:2019gzr}, etc.
There are also publications regarding the development of the event reconstruction based on the same algorithm principle designed for JUNO-type (spherical shape) liquid scintillator detectors~\cite{Liu:2018fpq, Wu:2018zwk, Li:2021oos, Huang:2021baf, Huang:2022zum}.
In comparison to the previous studies for the JUNO-type detector, the uniqueness of this algorithm is as follows:
\begin{itemize}
\item Using neutron samples, including fast-neutrons and $\gamma$-rays from a radioactive neutron source, as well as $\gamma$-rays from neutron captures, to prepare the input tables for the likelihood calculation (see Sections~\ref{subsec:simulation1} and~\ref{subsec:simulation2}).
\item PMT charge responses are modeled based on the laser calibration samples (see Section~\ref{subsec:simulation3}).
\item Identification of particle type in each event, separating either electrons/positrons/$\gamma$-rays from $\alpha$-particles/fast-neutrons, using pulse-shape discrimination techniques similar to that in Ref.~\cite{Ranucci:1998bc}.
\item The current JUNO standard event reconstruction algorithm, described in Refs.~\cite{Huang:2021baf, Huang:2022zum, JUNO:2024fdc}, requires $^{68}$Ge calibration samples collected at nearly 300 different points within the detector, which is expected to take over 30~hours to complete.
In contrast, the algorithm presented in this paper can be constructed using significantly fewer calibration samples, relying on neutron and laser calibrations, as well as continuously available cosmogenic neutron events, as presented in Section~\ref{sec:simulation}.
\end{itemize}
Furthermore, the development approach of the event reconstruction algorithm discussed in this paper can be directly applied to calibration data to be obtained from the actual detector. 
Compared to machine learning–based reconstruction methods, it does not rely on elaborate detector simulations and is therefore expected to be more readily applicable to early-stage physics analyses.
The rest of this paper is organized as follows.
A summary of the JUNO detector simulation and event samples generated for this study is presented in Section~\ref{sec:simulation}, before describing the details of the event reconstruction algorithm in Section~\ref{sec:eventreco}.
Section~\ref{sec:table} details the construction method for the look-up tables to be applied in the event reconstruction algorithm.
The performance of the algorithm estimated with the detector simulation is discussed in Section~\ref{sec:performance}, before the conclusion of the paper in Section~\ref{sec:conclusion}.

\section{Simulation of event samples}
\label{sec:simulation}
As the JUNO detector is currently under construction, this study was performed using the JUNO detector simulation software~\cite{Lin:2022htc}.
Particle tracking and generation/propagation of the scintillation and Cherenkov photons from the charged particles, within a detailed implementation of the JUNO's geometrical design~\cite{Li:2018fny}, are simulated with a customized Geant4-based software~\cite{GEANT4:2002zbu, Allison:2016lfl}.
Optical photons produced by charged particles experience various interaction processes including light absorption, scattering in the medium, refraction at the boundaries of the different detector media, and reflection on the surface of the detector components.
These light production and propagation processes in the liquid scintillator are modeled based on the real measurement in the Daya-Bay detector~\cite{JUNO:2020bcl}.
For the light emission time profile of scintillation photons, there are four components with different light emission decay times, and they are separately modeled for electrons/positrons, protons, and $\alpha$-particles.
For each optical photon reaching the PMT photocathode, detection is probabilistically determined based on the wavelength-dependent detection efficiency.
Individual PMT responses, including photon detection efficiency, charge amplification (and associated fluctuations), waveform generation, transit time (with its spread),  dark noise, and after-pulsing, as well as the electronics responses, are taken into account in the custom simulation package.
The parameters which quantify the responses of these detector instruments in this work are based on independent measurements~\cite{JUNO:2022hlz, Cao:2021wrq, Wang:2022tij, Cerrone:2022dvp, Coppi:2023nlv, Triozzi:2023zct, JUNO:2020orn}.
The total number of observed photoelectrons (p.e.) per unit energy estimated based on 2.2~MeV gamma-rays generated at the center of the detector is approximately 1,600~p.e./MeV. \par
There are two types of LPMTs in the JUNO detector: 4,997 of the 17,612 are Dynode LPMTs manufactured by Hamamatsu Photonics K.K. (model: R12860), while the remaining are Micro-Channel Plate PMTs (MCP-LPMTs) manufactured by North Night Vision Technology (model: N6201).
For the simulation of generated waveforms, the height of each pulse follows a combination of a Gaussian and an exponential distribution.
The component following the exponential function accounts for 1\% of the total waveforms for the Dynode LPMTs.
For the MCP LPMTs, a more prominent long-tail structure in the charge distribution (see Figure~\ref{fig:chargepdf1}) is observed~\cite{Zhang:2021ruy}, accounting for roughly 10\% of the total waveforms.
This is considered to occur because the photoelectrons emitted from the photocathode may hit the surface of the first electrode before entering the MCP channels, thereby producing additional secondary electrons~\cite{Weng:2024tjs}.
The SPMTs are Dynode PMTs produced by Hainan Zhanchuang Photonics Technology (model: XP72B22).
Table~\ref{tab:simulation1} summarizes the representative characteristics of these three types of PMTs. 
While the SPMTs offer better timing resolution compared to the LPMTs, their smaller photocathode area results in an insufficient number of hits from calibration events to construct a reliable time-response look-up table (timing PDF discussed in Section~\ref{subsec:timingpdf}). 
Therefore, they are not used for the vertex reconstruction and are instead utilized solely for the energy reconstruction.
In this paper, differences in individual PMT responses, such as gain, timing offsets, and charge non-uniformity, are assumed to have already been calibrated using methods described in previous studies~\cite{JUNO:2020xtj, Cabrera:2023dek}. \par

\begin{table}
\caption{
Representative characteristics (average values) of the PMT responses implemented in the JUNO detector simulation. 
Here, ``P.D.E.'', ``S.P.E. $\sigma$'', ``T.T.S. $\sigma$'', and ``D.N.R.'' represent photon detection efficiency at 420~nm of photon wavelength, single photoelectron charge resolution in a Gaussian component,  transit time spread, and dark noise rate, respectively.
As mentioned in the text, the terms ``LPMT'' and ``SPMT'' refer to the 20-inch large PMTs and 3-inch small PMTs, respectively.
}
\label{tab:simulation1}
\centering
\begin{tabular}{cccc}
\hline
 & Dynode LPMT & MCP LPMT & SPMT \\
\hline
\hline
P.D.E. & 30\% & 32\% & 25\% \\
\hline
S.P.E. $\sigma$ & 28\% & 33\% & 33\% \\
\hline
T.T.S. $\sigma$ & 1.2~nsec & 5.1~nsec & 0.7~nsec \\
\hline
D.N.R. & 9~kHz & 29~kHz & 0.5~kHz \\
\hline
\end{tabular}
\end{table}


In JUNO, the waveform at each LPMT is analyzed, and the observed charge and photon hit time are extracted by applying the waveform reconstruction.
This study employed the JUNO default waveform reconstruction, the deconvolution algorithm developed in the Daya-Bay experiment~\cite{Huang:2017abb}.
In this analysis, the peak value of the single photoelectron charge distribution is used as a conversion factor to express the observed charge in units of p.e.
In contrast to the LPMT system, waveform information is not saved in the SPMT system, and the pulse hit time, defined by the time when the waveform exceeds the constant threshold value, is saved.
In the present study, the SPMT output was processed with a digital counting method, i.e., only the hit information was considered in the energy reconstruction algorithm, instead of the observed charge at each SPMT. 
More details can be seen in Section~\ref{subsec:eventreco2}. \par
As already mentioned, the event reconstruction algorithm developed in this study can be tuned based on the detector calibration events.
Therefore, a variety of calibration events were generated, using the simulation package described above, where the waveform-reconstructed variables in each event were analyzed to build a set of look-up tables, which were then fed into the reconstruction algorithm.
Besides the calibration samples, uniformly distributed positron, $\alpha$-particle, and fast-neutron events were generated throughout the detector to evaluate the reconstruction performance.
The simulated event samples generated for this study are summarized in Table~\ref{tab:simulation2}, and the features of each calibration sample are described in the remainder of this section.

\begin{table*}
\caption{Summary of the simulation samples analyzed in this study.
In this table, $E_{\rm kin}$ stands for the initial kinetic energy of the simulated particle.
}
\label{tab:simulation2}
\centering
\begin{tabular}{llll}
\hline
Sample & Number of events & Event position & Purposes \\
\hline
\hline
Radioactive $^{241}$Am$^{13}$C  & 100k at each position & $X=0, Y=0,$ & Timing PDF construction \\
calibration source events & & and $Z=0, \pm6,$ & (Section~\ref{subsec:timingpdf}) \\
(Section~\ref{subsec:simulation1}) & & $\pm8, \pm10, \pm12, \pm13,$ & \\
 & & $\pm14, \pm15, \pm15.8$ & \\
 & & $\pm15.9, \pm16, \pm16.1$ & \\
 & & $\pm16.5, \pm16.8, \pm17$ & \\
 & & $\pm17.2, \pm17.5$~m & \\
\hline
Thermal neutrons & 1M & Uniformly distributed & Charge map construction \\
(Section~\ref{subsec:simulation2}) & & within the liquid & (Section~\ref{subsec:chargemap}) \\
 & & scintillator volume &  \\
\hline
Various intensity laser & 1M in total & Detector center & Charge PDF construction \\
(Section~\ref{subsec:simulation3}) & & $(X=0, Y=0, Z=0)$ & (Section~\ref{subsec:chargepdf}) \\
\hline
Positrons with $E_{\rm kin} =$ 0, & 100k for each $E_{\rm kin}$ value & Uniformly distributed & Performance check \\
0.5, 1, 2, 3, 4, 5, 8 11~MeV & & within the liquid & (Section~\ref{sec:performance}) \\
 & & scintillator volume & \\
\hline
$\alpha$ and fast-neutrons with $E_{\rm kin}$ & 100k for each particle & Uniformly distributed & Performance check \\
 ranging from 1 to 10~MeV & & within the liquid & (Section~\ref{sec:performance}) \\
 & & scintillator volume & \\
\hline
\end{tabular}
\end{table*}



\subsection{Radioactive $^{241}$Am$^{13}$C source}
\label{subsec:simulation1}
A radioactive americium-carbon ($^{241}$Am$^{13}$C) calibration source will be employed as a neutron source in the JUNO detector calibration program~\cite{Liu:2015cra, JUNO:2020xtj}.
Neutrons are produced in the following $\alpha$-n reactions:
\begin{align}
^{241}{\rm Am} &\rightarrow ^{237}{\rm Np}+\alpha, \label{eq:simulation4} \\
^{13}{\rm C}+\alpha &\rightarrow ^{16}{\rm O}^{(*)}+{\rm n}, \label{eq:simulation1} 
\end{align}
\begin{numcases}{^{16}{\rm O}^{*}\rightarrow}
^{16}{\rm O}+{\rm e^{+}}+{\rm e^{-}} & $({\rm 1st~excited~state})$, \label{eq:simulation2} \\
^{16}{\rm O}+\gamma & $({\rm 2nd~excited~state})$ \label{eq:simulation3}.
\end{numcases}

This source makes a characteristic delayed-coincidence event structure, similar to those produced by an inverse beta decay due to electron antineutrinos.
The prompt events in this calibration sample consist of the following four types of events (also summarized in Table~\ref{tab:simulation3}):
\begin{itemize}
\item Fast-neutron events from Reaction~(\ref{eq:simulation1}). 
About 84\% of the $\alpha$-n reactions leave the oxygen nucleus in the ground state, where the kinetic energy of the neutron ranges from 3 to 8~MeV. 
These fast-neutrons may elastically scatter off protons in the liquid scintillator, producing measurable scintillation photons.
\item $\gamma$-ray emission from $^{12}{\rm C}^{*}$. 
Fast-neutrons released by the $^{241}$Am$^{13}$C source may interact with $^{12}$C in the liquid scintillator via inelastic scattering, producing $^{12}{\rm C}^{*}$ which de-excites and emits a 4.4~MeV $\gamma$-ray.
\item Positron annihilation from Reaction~(\ref{eq:simulation2}).
Oxygen is produced in its first excited state from the $\alpha$-n interaction (in roughly 8\% cases), where an electron-positron pair is emitted upon its de-excitation.
However, the radioactive calibration source is made up of a source capsule made of stainless steel and Polytetrafluoroethylene (PTFE)~\cite{JUNO:2020xtj, Zhang:2021tob}. 
Both the electron and positron deposit energy and stop within the source capsule before reaching the liquid scintillator.
Additionally, the positron annihilates with an electron within the source, producing two 511~keV $\gamma$-rays, which can escape the source, depositing energy into the liquid scintillator, and giving off measurable scintillation photons.
\item $\gamma$-ray events from Reaction~(\ref{eq:simulation3}).
The remaining 8\% of the $\alpha$-n reactions leave $^{16}$O in its second excited state.
A 6.1~MeV $\gamma$-ray is emitted from the de-excitation process, which often escapes from the calibration source capsule and deposits its energy in the liquid scintillator.
\end{itemize}

\begin{table*}
\caption{
Summary of the observed prompt events from the radioactive americium-carbon ($^{241}$Am$^{13}$C) calibration source.
The primary $\alpha$-n reaction in the $^{241}$Am$^{13}$C source is described in Reactions~(\ref{eq:simulation4}),~(\ref{eq:simulation1}),~(\ref{eq:simulation2}) and~(\ref{eq:simulation3}) in the text.
}
\label{tab:simulation3}
\centering
\begin{tabular}{l|l|l|l}
\hline
Particle type & Kinetic energy & Production process & Probability \\
\hline
\hline
Neutron & 3 to 8~MeV & $\alpha$-n reaction & 84\% of the $\alpha$-n reactions \\
\hline
$\gamma$-ray & 4.4~MeV & $^{12}{\rm C}^{*}$ created by fast-neutrons in the liquid scintillator & 2\% of the fast-neutrons \\
\hline
Two $\gamma$-rays & 511~keV for each & Annihilation of $e^{+}$ emitted from $^{16}{\rm O}^*$ (the first excited state) & 8\% of the $\alpha$-n reactions \\
\hline
$\gamma$-rays & 6.1~MeV & From $^{16}{\rm O}^*$ (the second excited state) & 8\% of the $\alpha$-n reactions \\
\hline
\end{tabular}
\end{table*}

For the delayed events, as already mentioned above, neutrons from the $\alpha$-n reactions will thermalize in the liquid scintillator and be captured by either a hydrogen or $^{12}$C nucleus, resulting in the characteristic $\gamma$-ray emissions. \par
In the JUNO detector, one of the calibration source deployment systems, the Automatic Calibration Unit (ACU) system, can place the calibration source along the central axis ($Z$-axis) of the detector~\cite{JUNO:2020xtj, Hui:2021dnh}.  
To simulate this situation, radioactive $^{241}$Am$^{13}$C source simulation samples were generated along the central axis, in order to inspect the detector timing response for events at different locations.
The $^{241}$Am$^{13}$C source events provide a timing probability density function (PDF, implemented as a look-up table) for PMTs used for the vertex reconstruction as well as particle identification algorithms. 
Calibration using the ACU system is expected to be the most frequently performed among different calibration systems, and the timing PDF tables obtained from this calibration can be updated over time as the experiment progresses.
Note that the locations of the source shown in Table~\ref{tab:simulation2} were optimized based on the vertex reconstruction performance.
The details of the timing PDF construction are presented in Section~\ref{subsec:timingpdf}.

\subsection{Cosmic-ray induced spallation neutrons}
\label{subsec:simulation2}
Cosmic-ray muons passing through the JUNO detector produce neutrons through various spallation processes, on carbon nuclei in the liquid scintillator volume, and oxygen nuclei in the water volume outside the acrylic shell~\cite{Malgin:2017dwh, Wang:2001fq}.
Many liquid scintillator and water Cherenkov detectors have reported neutron tagging techniques and measurements of the neutron production yield~\cite{Hertenberger:1995ae, Boehm:2000ru, AberdeenTunnelExperiment:2015uaa, DayaBay:2017txw, KamLAND:2009zwo, LVD:1999ezh, Borexino:2013cke, Super-Kamiokande:2022cvw}.
Those cosmic-ray-induced spallation neutrons are expected to be distributed throughout the JUNO detector.
According to simulation studies, which accounted for the JUNO detector size, overburden, as well as cosmic-ray muon flux, the expected rate for the 2.2~MeV $\gamma$-ray event from the spallation neutron capture is roughly estimated to be 1.6~Hz, corresponding to approximately $\sim$one million events per week. \par
Neutrons produced via spallation are typically captured at an average distance of approximately one meter from the muon track, depending on their kinetic energy.
The selection of these spallation neutron events in the JUNO detector is currently under development, including muon track reconstruction~\cite{Genster:2018caz, Zhang:2018kag, Liu:2021okf, Yang:2022din}, and selections at the online and offline analysis levels~\cite{JUNO:2021vlw}.
In this study, uniformly distributed neutrons and their subsequent captures were simulated.
The $\gamma$-ray events produced by the neutron captures were exploited to make a table (charge map) that predicts the mean light luminosity at each PMT for a given event vertex position.
As previously mentioned, since spallation neutrons are continuously generated within the detector by cosmic-ray muons, the tables based on these events can also be updated during the experimental operation period.
The method to construct the charge map is presented in Section~\ref{subsec:chargemap}.

\subsection{Laser calibration}
\label{subsec:simulation3}
A laser calibration system is planned to be installed in the JUNO detector, with its basic design documented in Refs.~\cite{Zhang:2018yso, Takenaka:2024ctk}.
A laser diffuser ball will be attached to the tip of an optical fiber, isotropically emitting photons within the detector, and it will be deployed at the center of the detector with the ACU system.
The wavelength of the laser light is approximately 266~nm, where photons emitted from the diffuser ball are quickly absorbed by the liquid scintillator, which emits photons following the emission spectrum of the JUNO liquid scintillator~\cite{JUNO:2020bcl}.
The trigger signal of the laser calibration events will be externally generated, resulting in the high purity of these events.
The likelihood functions in the event reconstruction discussed in Section~\ref{sec:eventreco} require to capture the evolution of the LPMT charge distribution for different light luminosities at each LPMT.
Since the output light intensity of the laser device will be configurable and can cover more than 4 orders of magnitude, laser calibration events are suitable to produce LPMT charge response tables over a wide range of light intensities (charge PDF).
To demonstrate the methodology, in the detector simulation, 36 laser samples with different light intensities, ranging from single-photoelectron level to 200-photoelectron level at each LPMT, were generated, where the usage of these samples to prepare the charge PDF tables is detailed in Section~\ref{subsec:chargepdf}.

\section{Event reconstruction algorithm}
\label{sec:eventreco}
The algorithm consists of three sub-algorithms; event vertex reconstruction, energy reconstruction, and particle identification algorithms.
In this study, event vertex and energy reconstruction are performed separately, considering that the photoelectron detection timing information contributes little to the energy reconstruction and that increasing the number of fitting parameters significantly increases the computational cost.
The event vertex position is reconstructed prior to the other two algorithms which take the vertex position as the input.
All of the sub-algorithms execute likelihood calculations with the charge and photon hit times at individual LPMTs in each event. 
The energy reconstruction algorithm additionally includes the SPMT hit information.
The rest of this section describes the specific likelihood function for each algorithm.

\subsection{Event vertex position reconstruction}
\label{subsec:eventreco1}
There are two types of likelihood functions in the event vertex position reconstruction algorithm.
One primarily relies on LPMT timing information with fewer input tables, while the other exploits both charge and timing information observed in the LPMT system, achieving a better vertex reconstruction resolution.
The likelihood function for the first case is given as follows:
\begin{align}
L(t_{0}, \bm{x}) &= \prod_{i}^{\rm Hit~LPMTs}P^{t}_{i}(t_{i, {\rm res}}|R, \theta_{i, \rm PMT}, Q_{i, \rm obs}, Q_{\rm tot}), \label{eq:eventreco1}
\end{align}
where
\begin{align}
t_{i, {\rm res}} &= t_{i, {\rm first}} - {(\rm T.O.F.)}_{i} - t_{0}, \label{eq:eventreco2} \\
Q_{\rm tot} &= \sum_{i}^{\rm All~LPMTs}\left(Q_{i, \rm obs}-\mu_{i, \rm dark}\right).
\end{align}
Here, ``Hit LPMTs'' denote the LPMTs recording at least one hit, $R$ is the event radial position, $\theta_{i, \rm PMT}$ is the relative angle between the vertex position and the $i$th LPMT (see Figure~\ref{fig:detector}), $Q_{i, \rm obs}$ is the observed charge at the $i$th LPMT, $t_{i, {\rm first}}$ is the reconstructed first photoelectron hit time by the waveform decovolution algorithm at the $i$th LPMT in each event, $({\rm T.O.F.})_{i}$ stands for ``time-of-flight'' of the photon reaching from the vertex position to the $i$th LPMT, and $P^{t}_{i}$ represents the timing probability density function (timing PDF), reflecting the probability of observing a photon at the residual time of $t_{i, {\rm res}}$ with a given vertex position ($\bm{x}$) and event time $(t_{0})$ at the $i$th LPMT.
The parameter dependence of the timing PDF is described in Section~\ref{subsubsec:timingpdf1}.
The photon's time-of-flight is derived by dividing the distance of the light path from the vertex position to each LPMT by the effective speed of light in the detector medium:
\begin{align}
{(\rm T.O.F.)}_{i} &= (d_{i, \rm LS}\times n_{\rm LS}+d_{i, \rm water}\times n_{\rm water})/{\rm c}, \label{eq:eventreco3}
\end{align}
where
\begin{align}
n_{\rm LS} &= 1.55, \\
n_{\rm water} &= 1.38.
\end{align}
Here, $d_{i, \rm LS} (d_{i, \rm water})$ is the distance of the light path in the liquid scintillator (water) region from the vertex position to the $i$th LPMT position, ${\rm c}$ is the speed of light in vacuum, and $n_{\rm LS} (n_{\rm water})$ is the refractive index of the liquid scintillator (water). 
The light path calculation in the JUNO detector was discussed in Ref.~\cite{Liu:2018fpq} including the treatment for internal reflected photons, and the same method was adopted in this study.
$n_{\rm LS}$ and $n_{\rm water}$ are effective values and were chosen such that the LPMT photon hit times after subtracting T.O.F. are aligned for the $^{241}$Am$^{13}$C calibration events at different locations.
As for the timing PDF tables, they were obtained from the radioactive $^{241}$Am$^{13}$C source calibration events produced along the central axis of the detector, as introduced in Section~\ref{subsec:simulation1}.
Assuming rotational symmetry, the timing PDFs were tabulated as a function of the event radial position, relative angle between the event position and LPMT position ($\theta_{i, \rm PMT}$, see Figure~\ref{fig:detector}), observed charge, as well as total charge of the event $(Q_{\rm tot})$.
The total observed charge in each event is evaluated by summing up the observed charge at individual LPMTs and statistically subtracting the expected dark noise charge contributions $(\mu_{i, \rm dark})$ based on the dark noise rate at each LPMT (see also Equation~(\ref{eq:eventreco9})).
Details of the timing PDF construction are discussed in Section~\ref{subsec:timingpdf}.
The event vertex position is searched for by maximizing the likelihood function, i.e., scanning likelihood values as a function of vertex position.
In practice, this is done by minimizing the $(-\log L)$ value using the Minuit package~\cite{James:1975dr}. \par

The other likelihood function utilizes the LPMT charge information, in addition to the timing information, and is described by the following formula:
\begin{align}
L(t_{0}, \bm{x}) &= \prod_{j}^{\rm Unhit~LPMTs}P^{q}_{j}({\rm unhit}|\mu_{j, \rm exp}) \nonumber \\
&\times\prod_{i}^{\rm Hit~LPMTs}P^{q}_{i}(Q_{i, \rm obs}|\mu_{i, \rm exp}) \nonumber \\
&\times P^{t}_{i}(t_{i, {\rm res}}|R, \theta_{i, \rm PMT}, Q_{i, \rm obs}, Q_{\rm tot}), \label{eq:eventreco4} 
\end{align}
where
\begin{align}
\mu_{i, \rm exp} &= \mu_{i, \rm S}+\mu_{i, \rm dark} \nonumber \\
&=\mu_{{\rm S}, 0}(\bm{x}, \theta_{i, \rm PMT})\times Q_{\rm tot}+\mu_{i, \rm dark}, \label{eq:eventreco5} \\
\mu_{i, \rm dark} &= ({\rm D.N.R.})_{i}\times T_{\rm window}\times G_{i}. \label{eq:eventreco9}
\end{align}
Here, $P^{q}_{j}({\rm unhit}|\mu_{j, \rm exp})$ is the charge probability density function (charge PDF), reflecting the probability of not observing a photon with a given expected charge value $(\mu_{j, \rm exp})$ at the $j$th LPMT, and $P^{q}_{i}(Q_{i, \rm obs}|\mu_{i, \rm exp})$ is the probability of observing $Q_{i, \rm obs}$ with a given expected charge $(\mu_{i})$ at the $i$th LPMT.
The charge PDF look-up table is constructed using laser calibration samples of various light luminosities, where the construction method is detailed in Section~\ref{subsec:chargepdf}.
The expected charge at each LPMT is decomposed into a signal contribution due to scintillation and Cherenkov photons, along with a dark noise contribution.
The signal contribution is obtained from another look-up table $(\mu_{{\rm S}, 0}$, charge map), prepared as a function of the event vertex position $(\bm{x})$ and relative angle between the vertex and the LPMT ($\theta_{\rm PMT}$ in Figure~\ref{fig:detector}).
Meanwhile, the dark noise part $(\mu_{i, \rm dark})$ is calculated as the product of the dark noise rate $(({\rm D.N.R.})_{i})$, the timing window $(T_{\rm window} = 420~{\rm nsec})$, and the average charge per one PMT hit $(G_{i})$.
The parameter $(G_{i})$ is needed to convert the number of dark noise hits to its charge contribution.
As described in Section~\ref{subsec:chargemap}, the expected charge table is normalized by the total observed charge within $T_{\rm window}$ for each event and was built using the uniformly distributed neutron samples mentioned in Section~\ref{subsec:simulation2}. \par
Here, let us reiterate the three important look-up tables used in the likelihood functions as they are repeatedly discussed in this paper:
\begin{enumerate}
\item Timing PDF ($P^{t}(t_{{\rm res}}|R, \theta_{\rm PMT}, Q_{\rm obs}, Q_{\rm tot})$, see Equation~(\ref{eq:eventreco1})): the probability of observing the first photoelectron at the residual time of $t_{{\rm res}}$ with a given vertex position ($\bm{x}$) and event time $(t_{0})$ at each LPMT.
\item Charge map ($\mu_{{\rm S}, 0}(\bm{x}, \theta_{\rm PMT})$, see Equation~(\ref{eq:eventreco5})): the predicted light luminosity at each PMT as a function of the event position ($\bm{x}$) and angle to the PMT $(\theta_{\rm PMT})$, normalized by the total charge in each event.
Note that two other charge map tables normalized by the event energy ($\mu^{\prime}_{{\rm SPMT, S}, 0}({\bm x}, \theta_{\rm PMT})$, $\mu^{\prime}_{{\rm S}, 0}({\bm x}, \theta_{\rm PMT})$, see Equations~(\ref{eq:eventreco7}) and~(\ref{eq:eventreco8})) are introduced in the next section.
\item Charge PDF ($P^{q}(Q_{\rm obs}|\mu)$, see Equation~(\ref{eq:eventreco4})): the probability of observing $Q_{\rm obs}$ with a given expected charge $(\mu)$ at each LPMT.
\end{enumerate}

\subsection{Event energy reconstruction}
\label{subsec:eventreco2}
After the particle vertex position is determined, the event energy $(E_{\rm rec})$ in each event is estimated using the LPMT charge and SPMT hit information by maximizing the following likelihood function:
\begin{align}
L(E_{\rm rec}) &= \prod_{k}^{\rm Unhit~SPMTs}\exp\left(-\mu_{k,{\rm SPMT, exp}}\right) \nonumber \\
&\times\prod_{l}^{\rm Hit~SPMTs}\left(1-\exp\left(-\mu_{l,{\rm SPMT, exp}}\right)\right) \nonumber \\
&\times\prod_{j}^{\rm Unhit~LPMTs}P^{q}_{j}({\rm unhit}|\mu_{j, \rm exp}) \nonumber \\
&\times\prod_{i}^{\rm Hit~LPMTs}P^{q}_{i}(Q_{i, \rm obs}|\mu_{i, \rm exp}), \label{eq:eventreco6}
\end{align}
where
\begin{align}
\mu_{k, {\rm SPMT, exp}} &= \mu_{k, {\rm SPMT, S}}+\mu_{k, {\rm SPMT,dark}} \nonumber \\
&= \mu^{\prime}_{{\rm SPMT, S}, 0}({\bm x}, \theta_{k, \rm PMT})\times E_{\rm rec} \nonumber \\
& + \mu_{k, {\rm SPMT, dark}},  \label{eq:eventreco7} \\
\mu_{i,{\rm exp}} &= \mu_{i, {\rm S}}+\mu_{i, \rm dark} \nonumber \\
&= \mu^{\prime}_{{\rm S}, 0}({\bm x}, \theta_{i, \rm PMT})\times E_{\rm rec} + \mu_{i, \rm dark}. \label{eq:eventreco8}
\end{align}
Here, ``Hit SPMTs'' are defined by the SPMTs that record at least one hit within the event timing window $(T_{\rm window})$, and $\mu^{\prime}_{{\rm S}, 0}({\bm x}, \theta_{\rm PMT})$ $(\mu^{\prime}_{{\rm SPMT, S}, 0}({\bm x}, \theta_{\rm PMT}))$ represents the expected charge (mean light intensity) from the scintillation and Cherenkov photons, normalized by the event energy at the given reconstructed vertex position $({\bm x})$ and relative angle between the vertex position and each PMT $(\theta_{\rm PMT})$.
The charge map in the energy reconstruction is normalized by the $\gamma$-ray energy from neutron captures on hydrogen, i.e., 2.2~MeV, as these $\gamma$-ray events were used to build the charge map table in Section~\ref{subsec:chargemap}.
In the same manner as the likelihood function for the vertex reconstruction, the observed charge and expected charge at each LPMT are compared using the same charge PDF function, while only the hit information is included for the SPMTs, instead of the observed charge.
The unhit probability at each SPMT is computed by the Poisson probability, ${\rm Poisson}(0)=\exp{(-\mu)}$.
This treatment for the SPMT information is adopted in this study because it is anticipated that the absolute value of the observed charge at the SPMT is hard to calibrate in JUNO, due to the channel level dead time as well as threshold effects discussed in Ref.~\cite{JUNO:2020orn}.
For the reactor neutrino energy range, the light occupancy for the SPMTs is sufficiently low to ensure the validity of this method.


\subsection{Particle identification}
\label{subsec:eventreco3}
The algorithm presented in this paper can distinguish the primary particle in each event, separating electrons/positrons/$\gamma$-rays from $\alpha$-particles/fast-neutrons, by applying a pulse-shape discrimination technique to the observed timing distributions~\cite{Ranucci:1998bc}.
The timing PDF tables for $\gamma$-ray and fast-neutron events were separately prepared from the $^{241}$Am$^{13}$C calibration sources, discussed in Sections~\ref{subsec:simulation1} and~\ref{subsec:timingpdf}.
After reconstructing the event vertex position with the timing PDF tables constructed by the $\gamma$-ray samples, the timing-only likelihood presented in Equation~(\ref{eq:eventreco1}) is evaluated with another set of timing PDF tables, made based on the fast-neutron samples.
The difference between the two negative log-likelihood values measures which of the particle assumptions is favored.
In contrast to previous studies that derive the timing PDF from simulation samples~\cite{Cheng:2023zds}, this algorithm is expected to be more robust as it obtains the detector time response from calibration events.

\section{Look-up table preparation}
\label{sec:table}
This section describes the preparation of the input tables to be fed into the likelihood functions introduced in Section~\ref{sec:eventreco}, based on the event samples described in Section~\ref{sec:simulation}.

\subsection{Timing PDFs}
\label{subsec:timingpdf}
The timing PDF tables were made with the $^{241}$Am$^{13}$C calibration source events, located along the central axis of the detector. 
In each $^{241}$Am$^{13}$C source event, the residual time at each LPMT was calculated using Equations~(\ref{eq:eventreco2}) and~(\ref{eq:eventreco3}).
The residual time was computed by subtracting the photon time-of-flight (T.O.F.) and the event time $(t_{0})$ from the first photon hit at each LPMT in the $^{241}$Am$^{13}$C source event.
Here, the event time $t_{0}$ represents the event-by-event shift of the residual timing distribution relative to the trigger time, which is quantized with 16~nsec steps in JUNO.
To evaluate $t_{0}$ in each event, $(t_{\rm first}-{\rm T.O.F.})$ was calculated at each hit PMT and filled into a distribution.
The value of $t_{0}$ was obtained by searching for the peak position of this $(t_{\rm first}-{\rm T.O.F.})$ distribution.
By subtracting the obtained $t_{0}$ from $(t_{\rm first}-{\rm T.O.F.})$ in each event, the peak position of residual time $(t_{\rm first}-{\rm T.O.F.}-t_{0})$ was aligned among all of the collected $^{241}$Am$^{13}$C source events.
The calculation of T.O.F. was initially done based on the light propagation distance between the $^{241}$Am$^{13}$C source deployment position and LPMT position.
However, as pointed out in Ref.~\cite{Huang:2021baf}, neutrons and higher-energy $\gamma$-rays from the $^{241}$Am$^{13}$C source travel several tens of centimeters in the liquid scintillator volume, before being captured or depositing their energy, and the actual light emission point is away from the source location, causing a bias in the T.O.F. calculation.
Therefore, this calculation is modified in Section~\ref{subsubsec:timingpdf3} to mitigate such a bias. \par
\begin{figure}
\includegraphics[width=0.48\textwidth]{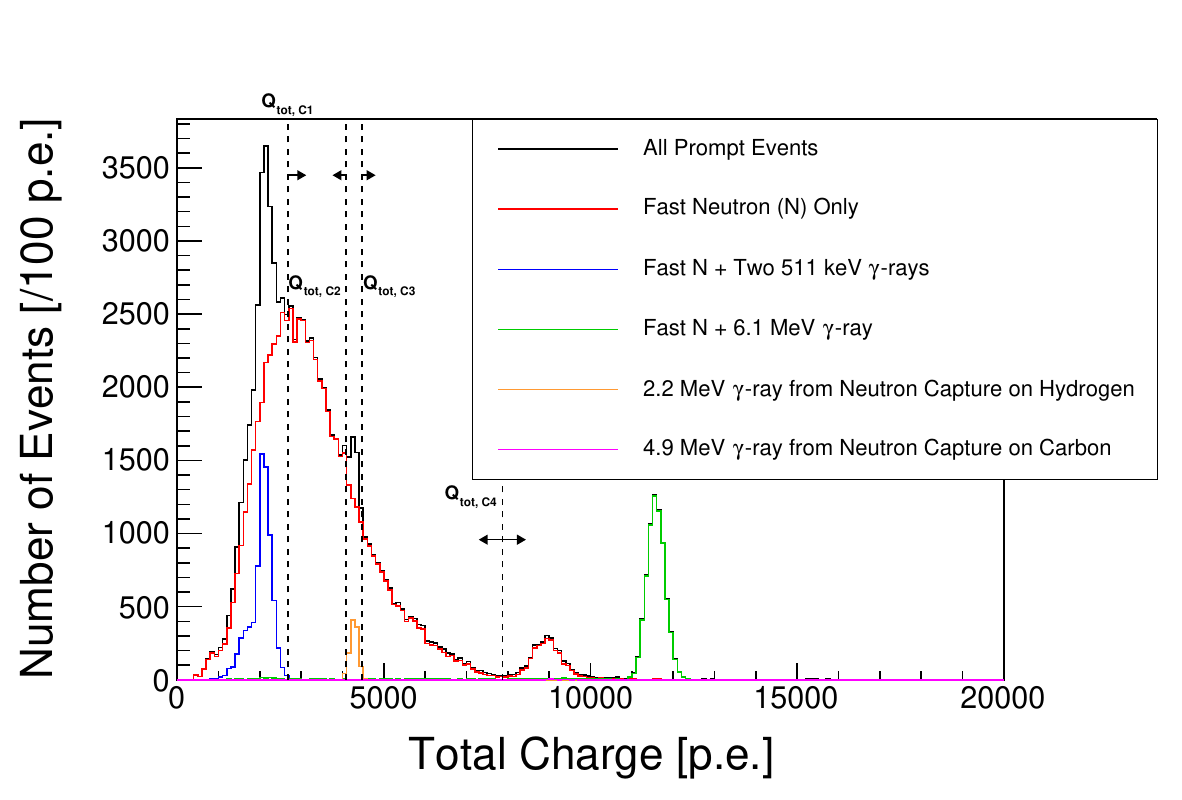}
\caption{
The total charge distributions for the prompt events in the $^{241}$Am$^{13}$C calibration source sample at the detector center.
The black histogram represents all of the $^{241}$Am$^{13}$C source events, and other colored histograms are made based on the true particle information from the simulation.
The vertical lines denote the event classification criteria as described in the text.
The peak around 9000~photoelectrons (p.e.) corresponds to the 4.4~MeV $\gamma$-rays from $^{12}{\rm C}^{*}$ produced by the inelastic scattering between fast-neutrons and $^{12}{\rm C}$ in the liquid scintillator volume.
A small fraction of 2.2~MeV $\gamma$-ray events contaminate this prompt event sample following the event time cut, yielding a peak around 4000~p.e.
}
\label{fig:timingpdf1}
\end{figure}

As introduced in Section~\ref{subsec:simulation1}, the $^{241}$Am$^{13}$C calibration source events consist of different types of events, among which 2.2~MeV $\gamma$-ray (delayed events), higher-energy $\gamma$-ray (prompt events), and fast-neutron (prompt events caused by pure proton-recoils) events were selected to build the timing PDF tables separately.
Using the delayed neutron capture time structure, prompt events and delayed events were separated with an event time cut:
\begin{align}
T_{\rm diff} > 1000~\mu{\rm sec} &\rightarrow \text{prompt event}, \nonumber \\
20 < T_{\rm diff} < 600~\mu{\rm sec} &\rightarrow \text{delayed event}, \nonumber
\end{align}
where $T_{\rm diff}$ is defined in each event by the elapsed time since the last triggered event.
The lower bound of the delayed event cut was set to eliminate spurious events induced by the PMT after-pulses~\cite{Zhao:2022gks}.
In addition, the higher-energy $\gamma$-ray and fast-neutron events were classified by the total observed charge in the event as shown in Figure~\ref{fig:timingpdf1}. 
The classification boundaries in the total charge distribution were determined based on the peak locations associated with  $2\times511$~keV $\gamma$-ray events, 2.2~MeV $\gamma$-ray events, and 4.4~MeV $\gamma$-ray events. 
For the prompt events, events with total charge in the regions from $Q_{\rm tot,C1}$ to $Q_{\rm tot,C2}$ and from $Q_{\rm tot,C3}$ to $Q_{\rm tot,C4}$ (shown in Figure~\ref{fig:timingpdf1}) are categorized as fast-neutron events.
Meanwhile, events with total charge higher than $Q_{\rm tot,C4}$ are defined as the higher-energy $\gamma$-ray events, composed of the 4.4~MeV $\gamma$-ray events from $^{12}{\rm C}^{*}$ and 6.1 MeV $\gamma$-ray events from $^{16}{\rm O}^{*}$, as introduced in Section~\ref{subsec:simulation1}. \par

\subsubsection{Parameter dependence}
\label{subsubsec:timingpdf1}
The probability density function for the residual time (timing PDF) observed in the LPMTs depends on the spatial relationship between the event position and PMT, observed charge, and event energy.
Therefore, the timing PDF tables were tabulated (binned) as a function of the event radial position ($R$), the relative angle between the event and LPMT positions $(\theta_{\rm PMT}$, see Figure~\ref{fig:detector}), observed charge value ($Q_{\rm obs}$), and total charge ($Q_{\rm tot}$) in each event.
Assuming rotational symmetry, the $R$ dependence was scanned by the $^{241}$Am$^{13}$C calibration source events located along the central axis, as shown in Table~\ref{tab:simulation2}.
The locations of the LPMTs are distributed at 122 different $\theta_{\rm PMT}$ values for events on the central axis of the JUNO detector, utilized to inspect the $\theta_{\rm PMT}$ dependence. 
As the observed charge $(Q_{\rm obs})$ increases, the first hit time $t_{\rm first}$ tends to be sooner and the shape of the distribution becomes narrower.
This effect is considered by separating the timing PDF tables using the observed charge value.
The width of the distribution is narrower for events with higher total charge (i.e., more PMT hits), as a greater number of PMTs detect photons that do not undergo optical scattering or reflection as their first hit.
Additionally, the event time ($t_0$) can be determined with higher precision in such events.
To properly take this into account, we separately prepared the timing PDF tables with 2.2~MeV $\gamma$-ray and higher-energy $\gamma$-ray samples, treating them as two bins using the mean total charge in each sample as a representative value.
Besides the event energy (total charge) dependence, the dependence of the timing PDF on the particle type is taken into account by separately preparing the timing PDF tables with the fast-neutron samples.
It should be noted here that the timing PDF tables made from the fast-neutron sample are looked up only when the particle identification algorithm is executed, while the vertex reconstruction looks up the timing PDF tables produced by the two different $\gamma$-ray samples (2.2~MeV and higher-energy), regardless of the true particle type in the event.
As shown in Table~\ref{tab:simulation1}, the timing resolution of the LPMT is quite different between MCP LPMTs and Dynode LPMTs.
Therefore, the timing PDF tables were separately prepared for each LPMT type.

\subsubsection{Construction of timing PDFs}
\label{subsubsec:timingpdf2}
\begin{figure}
\includegraphics[width=0.48\textwidth]{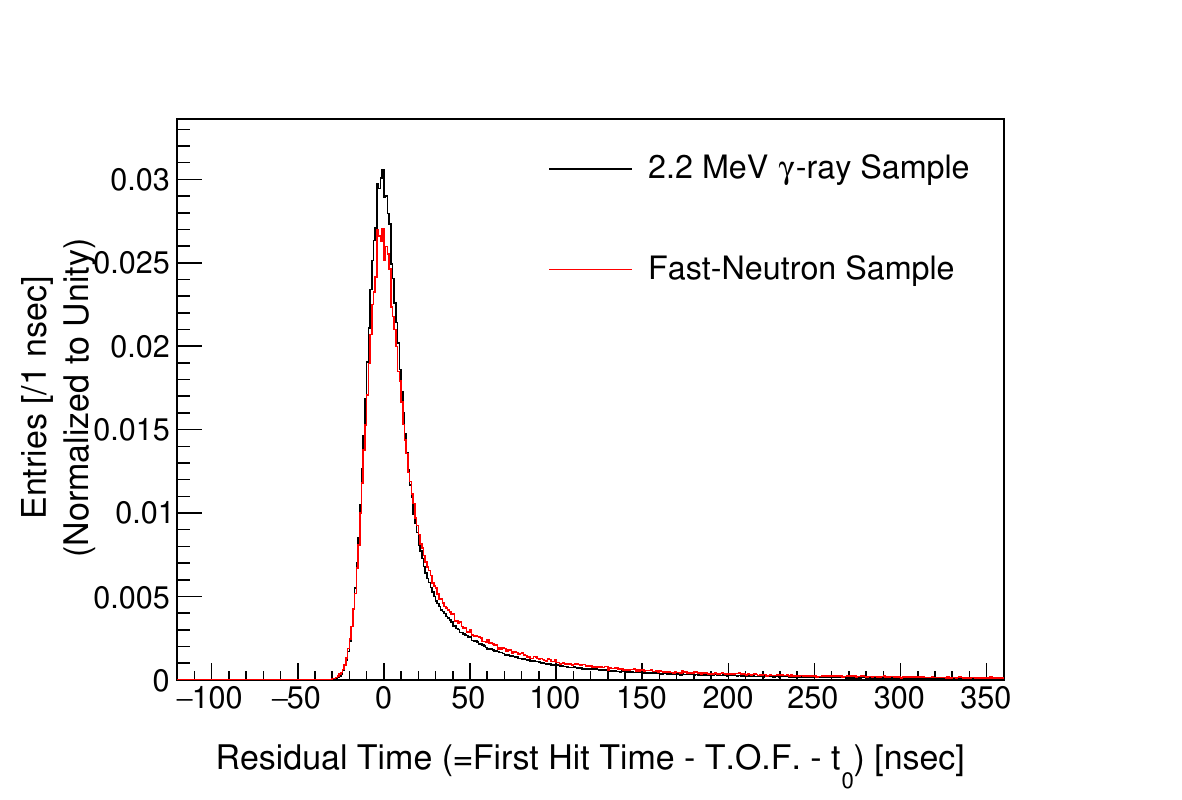}
\caption{
Example residual timing distributions for the MCP LPMTs in the $^{241}$Am$^{13}$C source events at $Z=10$~m.
Both of the histograms are normalized by their area.
The histogram from the fast-neutron sample has a longer tail than the one from the 2.2~MeV $\gamma$-ray sample.
}
\label{fig:timingpdf2}
\end{figure}

Figure~\ref{fig:timingpdf2} shows the typical residual timing distributions from the 2.2~MeV $\gamma$-ray and fast-neutron samples.
The light emission time profile of organic scintillators consists of multiple components, classified by their light emission decay times.
Compared to electrons and positrons (including ones created by $\gamma$-rays), heavier particles, such as $\alpha$-particles, and protons, including those scattered by fast-neutrons, deposit more energy per unit distance traveled (higher $dE/dx$).
This increased ionization power leads to a quenching effect, which reduces the yield of fast light emissions, relative to the slow light emission components~\cite{Birks:1964zz, Leo:1987kd}.
Hence, a more prominent long-tail structure can be seen in the residual time distribution for the fast-neutron sample, which helps to separate electron/positron/$\gamma$-ray events from those caused by heavier particles, $\alpha$-particles/fast-neutrons. \par

These residual time distributions were fitted with the following two exponentially modified Gaussian functions:
\begin{align}
f_{\rm S+D}(t) &= Cf_{\rm S}(t) + f_{\rm D}(t), \label{eq:timingpdf1}
\end{align}
where
\begin{align}
f_{\rm S}(t) &= {\bigg \{}\frac{\alpha}{2\tau_{1}}\exp\left(\frac{2\mu+\sigma^{2}/\tau_{1}-2t}{2\tau_{1}}\right) \notag \\
&\times{\rm erfc}\left(\frac{\mu+\sigma^{2}/\tau_{1}-t}{\sqrt{2}\sigma}\right) \notag \\
&+ \frac{(1-\alpha)}{2\tau_{2}}\exp\left(\frac{2\mu+\sigma^{2}/\tau_{2}-2t}{2\tau_{2}}\right) \notag \\
&\times{\rm erfc}\left(\frac{\mu+\sigma^{2}/\tau_{2}-t}{\sqrt{2}\sigma}\right){\bigg \}}, \label{eq:timingpdf2} \\
f_{\rm D}(t) &= (bkg_{1}-bkg_{2})\times\left(1-\int_{t_{\rm start}}^{t}f_{\rm S}(t^{\prime})dt^{\prime}\right) \notag \\
&+bkg_{2}, \label{eq:timingpdf3}
\end{align}
and $C$, $\alpha$, $\mu$, $\sigma$, $\tau_{1,2}$, and $bkg_{1,2}$ are fitting parameters.
Here, $t_{\rm start}$ is the lower end of the fitting time region and fixed to be -30~nsec, and ${\rm erfc}$ denotes the complementary error function.
Equation~(\ref{eq:timingpdf2}) fits the contributions from scintillation and Cherenkov photons from the calibration source events, which is modeled by a convolution of two exponential functions (fast and slow components) and a Gaussian function (PMT response, transit time spread).
Equation~(\ref{eq:timingpdf3}) represents the contributions from the dark noise hits.
As the timing PDF takes only the first hit time at each LPMT in the event timing window, the contributions from the dark noise hits compete with the scintillation and Cherenkov hits, leading to the dependence on the integral of $f_{S}(t)$ in Equation~(\ref{eq:timingpdf3}).
The fitting parameters $C, bkg_{1}$, and $bkg_{2}$ describe the scaling of the scintillation photons and dark noise contributions, while the parameters $\alpha$, $\mu$, $\sigma$, $\tau_{1}$, and $\tau_{2}$ describe the function shape.
This fitting was applied to the residual time distributions made from the $^{241}$Am$^{13}$C source events at different detector locations and different LPMT positions.
In the vertex reconstruction stage, the timing PDF value in Equation~(\ref{eq:eventreco1}) at each LPMT is calculated with a given observed residual time at the LPMT $(t_{\rm res, 0})$, test event radial position $(R)$, relative angle $(\theta_{\rm PMT})$, observed charge $(Q_{\rm obs})$, and total charge in the events $(Q_{\rm tot})$, by referring to the look-up tables prepared in this procedure. 
The parameter space between neighboring bins is linearly interpolated during the computation.
Although the values of $C$ (contribution from scintillation/Cherenkov photons) and $bkg_{1,2}$ (contribution from dark noise hits) in Equation~(\ref{eq:timingpdf1}) are obtained for 2.2~MeV and higher-energy $\gamma$-ray samples, their actual values at the event reconstruction stage are scaled according to the total charge in the event and dark noise rate at the LPMT.

\subsubsection{Iterative timing PDF construction}
\label{subsubsec:timingpdf3}
\begin{figure}
\includegraphics[width=0.48\textwidth]{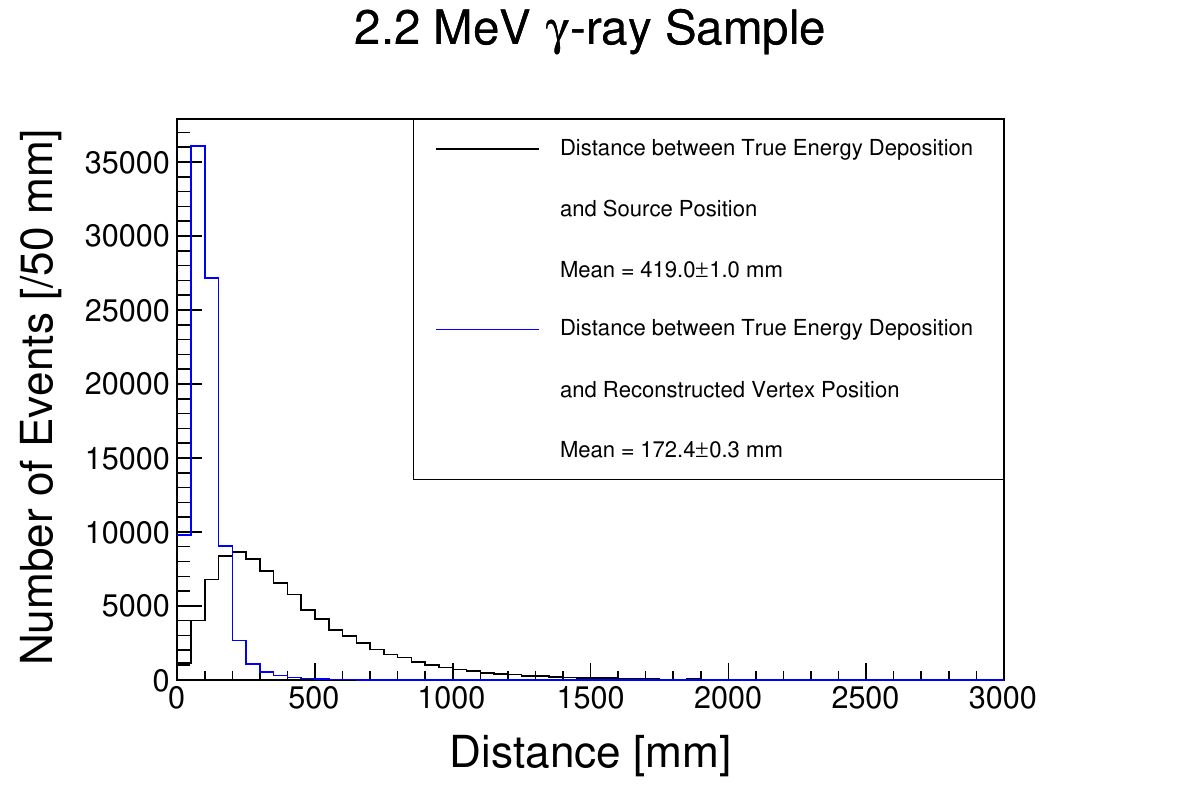}
\caption{
The black histogram shows the distance between the 2.2~MeV $\gamma$-ray true energy deposition position and $^{241}$Am$^{13}$C source position including the neutron and 2.2~MeV $\gamma$-ray travel length.
The blue histogram shows the distance between the true energy deposition and reconstructed vertex positions based on the vertex reconstruction algorithm before applying the iterative timing PDF construction introduced in Section~\ref{subsubsec:timingpdf3}.
}
\label{fig:timingpdf3}
\end{figure}

\begin{figure}
\includegraphics[width=0.48\textwidth]{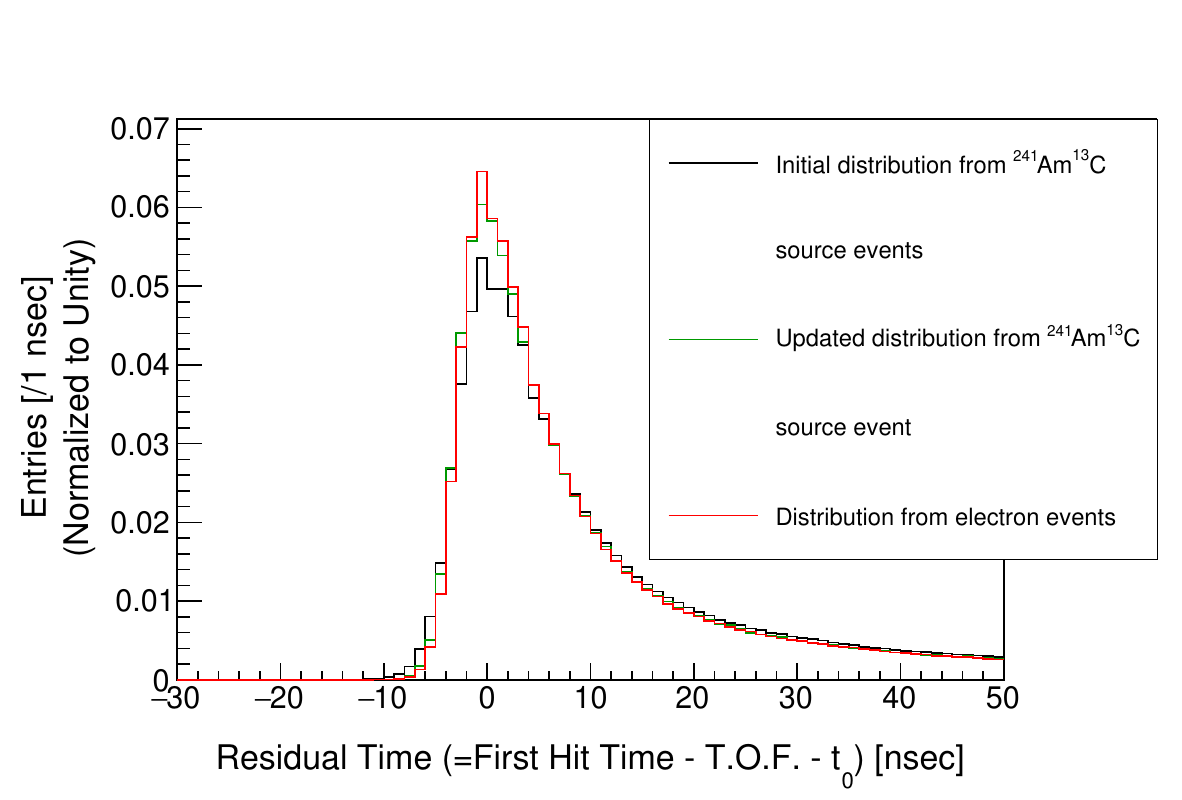}
\caption{
The residual time distributions of the $^{241}$Am$^{13}$C source events at (0, 0, 16~m) for the Dynode LPMTs.
The black histogram shows the original residual time distributions, where the photons' time-of-flight are calculated based on the $^{241}$Am$^{13}$C source position.
The green histogram corresponds to the residual time distribution with the iterative method introduced in the text.
The red distribution shows the residual time distribution from the electron sample generated at (0, 0, 16~m).
All of the histograms are normalized by their area.
}
\label{fig:timingpdf4}
\end{figure}

As introduced above, although each $^{241}$Am$^{13}$C source sample is associated with a fixed $R$ value, the fast-neutrons, $\gamma$-rays from the neutron captures, as well as higher-energy $\gamma$-rays from the $^{241}$Am$^{13}$C source, typically travel several tens of centimeters in the liquid scintillator before depositing their energy, as shown in the black histogram in Figure~\ref{fig:timingpdf3}.
However, the initial timing PDF construction assumes that scintillation and Cherenkov photons come from the source location in the time-of-flight calculation in Equation~(\ref{eq:eventreco2}), even though the actual light emission position is away from the source location.
This inconsistency between the actual and assumed light emission locations introduces an additional smearing in the obtained residual timing distribution from the $^{241}$Am$^{13}$C source events.
To highlight these effects, an electron sample with a kinetic energy of 2.2~MeV was produced at the same positions as the $^{241}$Am$^{13}$C source deployment positions, and the residual time distributions of the two samples are compared in Figure~\ref{fig:timingpdf4}.
Since electrons in this energy range travel only a few millimeters inside the liquid scintillator, the electron sample does not have the aforementioned issues, providing a more precise residual timing distribution at the given location. \par

To overcome this issue, the following additional treatments were adopted.
The event-by-event vertex position reconstructed by the algorithm shown in Equation~(\ref{eq:eventreco1}) with the initial time PDF tables was found to be a better approximation of the true vertex position compared to the fixed $^{241}$Am$^{13}$C source position as shown in the blue histogram in Figure~\ref{fig:timingpdf3}.
Therefore, the time-of-flight calculation in Equation~(\ref{eq:eventreco2}) was redone based on the event-by-event reconstructed vertex position, corrected by an observed average reconstruction bias along the radial direction (up to 10~cm) according to the source position.
In addition, events whose corrected-reconstructed vertex position was within 20~cm from the source location were selected, and only the selected events were used to build the time PDF tables to more precisely extract the detector response around the source location.
The green histogram in Figure~\ref{fig:timingpdf4} shows the residual time distribution made by applying the updated time-of-flight calculation to these selected events, approaching the one produced by the electron sample.
These procedures can be iteratively applied each time the timing PDF tables and vertex reconstruction are updated.
In this study, we found that two iterations were sufficient.
The vertex reconstruction performance with these improvements on the timing PDF tables is presented in Section~\ref{subsec:performance1}.

\subsection{Charge map}
\label{subsec:chargemap}
The likelihood calculations presented in Equations~(\ref{eq:eventreco4}) and~(\ref{eq:eventreco6}) require the evaluation of the expected light intensity ($\mu_{\rm exp}$) at each PMT with a given event vertex and energy.
This study employed the 2.2~MeV $\gamma$-ray sample from the spallation neutrons, as introduced in Section~\ref{subsec:simulation2}, and prepared the normalized expected charge tables (charge map), $\mu_{\rm S, 0}(\bf{x}, \theta_{\rm PMT})$, $\mu^{\prime}_{\rm SPMT, S, 0}(\bf{x}, \theta_{\rm PMT})$, and $\mu^{\prime}_{\rm S, 0}(\bf{x}, \theta_{\rm PMT})$ in Equations~(\ref{eq:eventreco5}),~(\ref{eq:eventreco7}) and~(\ref{eq:eventreco8}), respectively.
These charge map tables are normalized by the total charge or event energy (2.2~MeV) in the event, and they can be applied to events of other energies by scaling according to the total charge and energy in each event.

\subsubsection{Detector divisions}
\label{subsubsec:chargemap1}
Considering various light transportation processes, such as light absorption, scattering, refraction, and reflection, the light intensity reaching each PMT depends on the event vertex position, as well as the relative spatial relationship between them.
Hence, the charge map tables were constructed by dividing the liquid scintillator volume into many small regions (voxels), and the average charge value in each voxel was evaluated using the 2.2~MeV $\gamma$-ray events from neutron captures on hydrogen.
Assuming the azimuthal symmetry, the size and location of each small voxel are characterized by the cubic of the event radial position $(R^{3}{\rm , 74~bins})$, cosine of the event zenith angle $(\cos\Theta{\rm, 20~bins})$, and cosine of the relative angle to the PMT $(\cos\theta_{\rm PMT}{\rm, 162~bins})$ as follows (also see Figure~\ref{fig:detector} for their definition):
\begin{align}
\Delta R^{3} &= 
\begin{cases}
25~{\rm m}^{3} & (R^{3}<600~{\rm m}^{3}), \\
50~{\rm m}^{3} & (R^{3}>600~{\rm m}^{3}), \\
\end{cases} \label{eq:chargemap1} \\
\Delta\cos\Theta &= 0.1, \label{eq:chargemap2} \\
\Delta\cos\theta_{\rm PMT} &= 
\begin{cases}
2/960 & (\cos\theta_{\rm PMT}>0.9), \\
2/120 & (\cos\theta_{\rm PMT}<0.9).
\end{cases} \label{eq:chargemap3}
\end{align}
More bins are allocated in the detector regions where there is a large change in the average charge versus position, while fewer bins are introduced where the change of the average charge value is small.
These detector divisions were optimized to minimize the non-uniformity of the reconstructed energy to be presented in Section~\ref{subsec:performance2}.
The same type of PMTs (either MCP/Dynode LPMT or SPMT) sharing the same region $(R^{3}, \cos\Theta, \cos\theta_{\rm PMT})$ are merged in the charge map calculation.

\subsubsection{Charge map construction}
\label{subsubsec:chargemap2}
Contrary to the radioactive calibration sources deployed by the JUNO calibration source deployment systems, the event positions of the 2.2~MeV $\gamma$-rays from spallation neutrons are randomly distributed in the detector.
However, the vertex position of each 2.2~MeV $\gamma$-ray event has to be identified beforehand to make charge map tables, since they are parameterized as a function of event position and angle to PMT.
Each 2.2~MeV $\gamma$-ray event position was first reconstructed with the timing-only event reconstruction algorithm (Equation~(\ref{eq:eventreco1})).
This algorithm runs without relying on the charge map tables from the randomly distributed 2.2~MeV $\gamma$-ray events, and the event-by-event reconstructed vertex position was used as an input parameter for the charge map construction. \par 
The normalized average charge value for the LPMTs $(\mu_{{\rm S}, 0}, \mu^{ \prime}_{{\rm S}, 0})$ in each $(R^{3}, \cos\Theta, \cos\theta_{\rm PMT})$ voxel was computed using the following formula:
\begin{align}
\mu_{{\rm S}, 0} &= \frac{\sum_{k}^{N_{\rm Ev}}\left(\frac{\sum_{i}^{{\rm LPMTs~in~voxel}}\left(Q_{i, k}-\mu_{i, \rm dark}\right)}{Q_{k, {\rm tot}}}\right)}{\sum_{k}^{N_{\rm Ev}}N_{k,{\rm LPMTs~in~voxel}}}, \label{eq:chargemap5} \\
\mu^{\prime}_{{\rm S}, 0}&=\frac{\sum_{k}^{N_{\rm Ev}}\left(\frac{\sum_{i}^{{\rm LPMTs~in~voxel}}\left(Q_{i}-\mu_{i, \rm dark}\right)}{2.2}\right)}{\sum_{k}^{N_{\rm Ev}}N_{k,{\rm LPMTs~in~voxel}}}, \label{eq:chargemap6}
\end{align}
where
\begin{align}
Q_{k,\rm tot} &= \sum_{i}^{\rm {All~LPMTs}}(Q_{i, k}-\mu_{i, {\rm dark}}), \\
\mu_{i, \rm dark} &= ({\rm D.N.R.})_{i}\times T_{\rm window}\times G_{i}.  \label{eq:chargemap4}
\end{align}
Here, $Q_{i, k}$ is the observed charge at the $i$th LPMT in a given voxel in the $k$th event ($Q_{i, k}$ is 0 for non-fired LPMTs), $N_{k,{\rm LPMTs~in~voxel}}$ is the number of LPMTs (either MCP or Dynode LPMTs) falling into the same $(R^{3}, \cos\Theta, \cos\theta_{\rm PMT})$ voxel in the $k$th event, and $N_{\rm Ev}$ is the number of events in this voxel.
The average dark noise contribution $(\mu_{i, {\rm dark}})$ at each LPMT was calculated based on the dark noise rate, the length of the timing window of 420~nsec, and the average charge per PMT hit. \par
As for the charge map tables for the SPMTs, the mean light luminosity at each SPMT from scintillation and Cherenkov photons ($\mu^{\prime}_{{\rm SPMT, S, 0}}$) was evaluated with the SPMT hit information based on the Poisson statistics.
The number of PMT hits $(N_{\rm fired})$ out of a certain number of events $(N_{\rm total})$ can be generically associated with the mean light intensity $(\mu)$ using the zeroth term of the Poisson statistics:
\begin{align}
1-\frac{N_{\rm fired}}{N_{\rm total}} &= {\rm Poisson}(0, \mu) \nonumber \\
&= \exp{(-\mu)}. \label{eq:chargemap7}
\end{align}
Here, $\mu$ can be decomposed into a signal ($\mu_{\rm S}$) and dark noise ($\mu_{\rm dark}$) contributions as follows:
\begin{align}
\mu &= \mu_{\rm S} + \mu_{\rm dark}. \label{eq:chargemap8}
\end{align}
In a similar manner, the number of PMT hits caused by the dark noise $(N_{\rm dark})$ can expressed as follows:
\begin{align}
1-\frac{N_{\rm dark}}{N_{\rm total}} &= {\rm Poisson}(0, \mu_{\rm dark}) \nonumber \\
&= \exp{(-\mu_{\rm dark})}. \label{eq:chargemap9}
\end{align}
From Equations~(\ref{eq:chargemap7}),~(\ref{eq:chargemap8}), and~(\ref{eq:chargemap9}), $\mu_{\rm S}$ can be derived from the number of PMT hits and its dark noise contribution:
\begin{align}
\mu_{\rm S} &= -\log\left(\frac{N_{\rm total}-N_{\rm fired}}{N_{\rm total}-N_{\rm dark}}\right).
\end{align}

This formula can be extended as follows to calculate the normalized light intensity for the SPMTs at each voxel:
\begin{align}
&\mu^{\prime}_{\rm SPMT, S, 0} \nonumber \\
&=\frac{-\log\left(\frac{\sum_{k}^{N_{\rm Ev}}\left(N_{k, \rm SPMTs~in~voxel}-N_{k, \rm fired~in~voxel}\right)}{\sum_{k}^{N_{k, \rm Ev}}\left(N_{k, \rm SPMTs~in~voxel}-N_{k, \rm dark~in~voxel}\right)}\right)}{2.2}, 
\end{align}
where
\begin{align}
N_{\rm dark~in~voxel}&=\sum_{i}^{{\rm {SPMTs~in~voxel}}}({\rm D.N.R.})_{i}\times T_{\rm window}.
\end{align}
In each event, $N_{\rm SPMTs~in~voxel}$ is the number of SPMTs falling into the same $(R^{3}, \cos\Theta, \cos\theta_{\rm PMT})$ voxel, $N_{\rm fired~in~voxel}$ is the number of fired SPMTs in the same voxel, and $N_{\rm dark~SPMTs~in~voxel}$ is the expected number of dark noise hits from the SPMTs in the given voxel, which is also calculated based on the individual SPMT dark noise rate. \par
\begin{figure*}
\begin{minipage}{0.49\linewidth}
\centering
\includegraphics[width=1\linewidth]{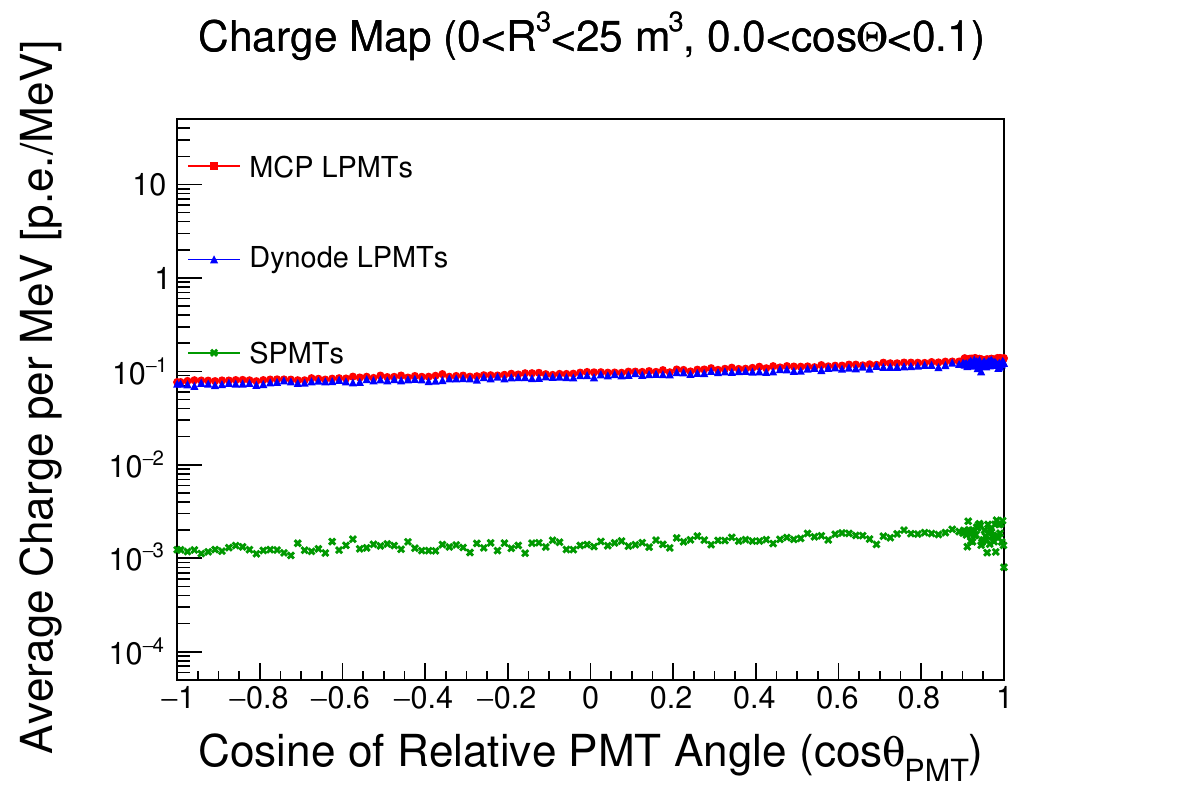}
\end{minipage}
\begin{minipage}{0.49\linewidth}
\centering
\includegraphics[width=1\linewidth]{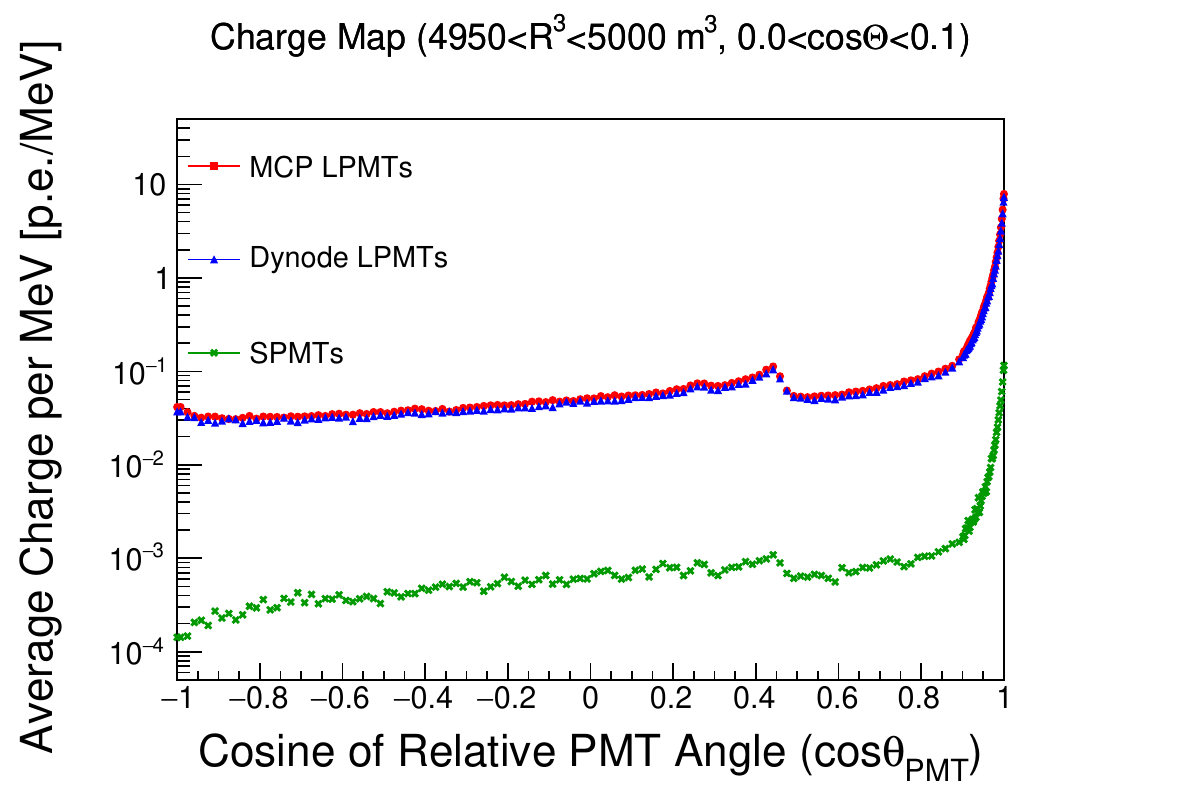}
\end{minipage}
\caption{
The left (right) plot shows the charge map distribution for events around the detector center (edge).
The red filled circles (blue triangles) represent the MCP (Dynode) LPMTs, and the green crosses show the SPMTs.
As expressed in Equation~(\ref{eq:chargemap3}), the bin width is not uniform along $\cos\theta_{\rm PMT}$ to capture the sharp change of the average charge, especially in the detector edges.
The lower charge scale for the SPMTs is due to their smaller light acceptance.
}
\label{fig:chargemap1}
\end{figure*}

Using one million 2.2~MeV $\gamma$-ray events from neutron captures on hydrogen, these computations were done for both the LPMTs and SPMTs at each detector voxel.
Example plots showing the charge map normalized by the $\gamma$-ray energy (2.2~MeV) at two different $R^{3}$ slices are displayed in Figure~\ref{fig:chargemap1}.
The average charge for the MCP LPMTs is systematically higher than that for the Dynode LPMTs, since the MCP LPMTs have a more significant tail in their charge distribution, as can be seen in Figure~\ref{fig:chargepdf1}.
Since the light acceptance of the SPMTs is roughly 50~times lower than the LPMTs, the vertical axis scale is also about 50~times smaller, though they share a similar distribution shape.
The small peak structures seen in the right plot of Figure~\ref{fig:chargemap1} are due to the total internal light reflection inside the detector, caused by the mismatch of the refractive index among the liquid scintillator, acrylic vessel, and water.
The PMTs sitting in the parameter space with $\cos\theta_{\rm PMT}$ ranging from 0.5 to 0.9 are substantially shadowed by this total internal reflection and do not receive direct photons from the particle, resulting in lower average charge values.

\subsection{Charge PDF}
\label{subsec:chargepdf}
After the expected charge is evaluated at each LPMT, based on the given event assumption (vertex position or energy) in the likelihood calculation, the probability of observing $Q$ with the given expected charge $\mu$ is computed using the charge probability density function (charge PDF) tables for the LPMTs.
This table was prepared for individual LPMTs and obtained using the various-intensity laser calibration samples described in Section~\ref{subsec:simulation3}.
The charge PDF tables are divided into two parts, the low-charge and high-charge.
This section will introduce the motivation for the charge PDF table construction with laser samples, followed by descriptions of their construction.

\subsubsection{Motivation}
\label{subsubsec:chargepdf1}
\begin{figure*}
\begin{minipage}{0.49\linewidth}
\centering
\includegraphics[width=1\linewidth]{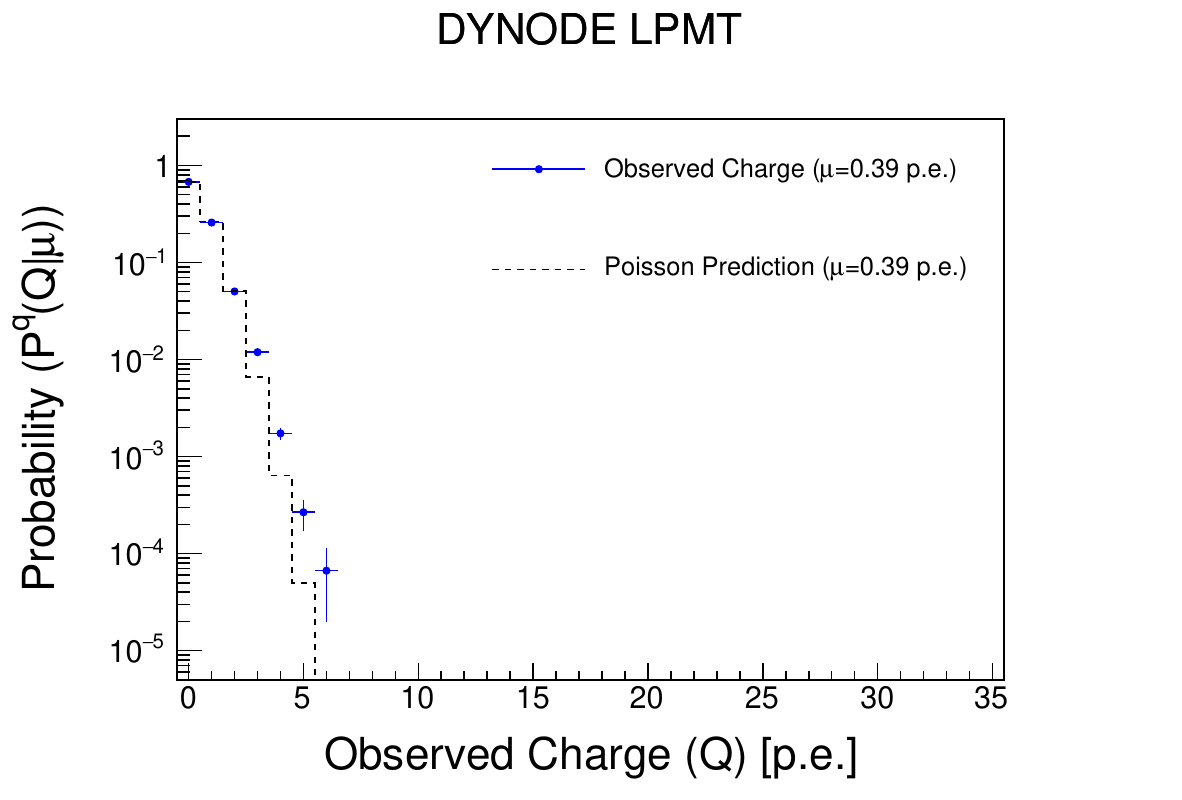}
\end{minipage}
\begin{minipage}{0.49\linewidth}
\centering
\includegraphics[width=1\linewidth]{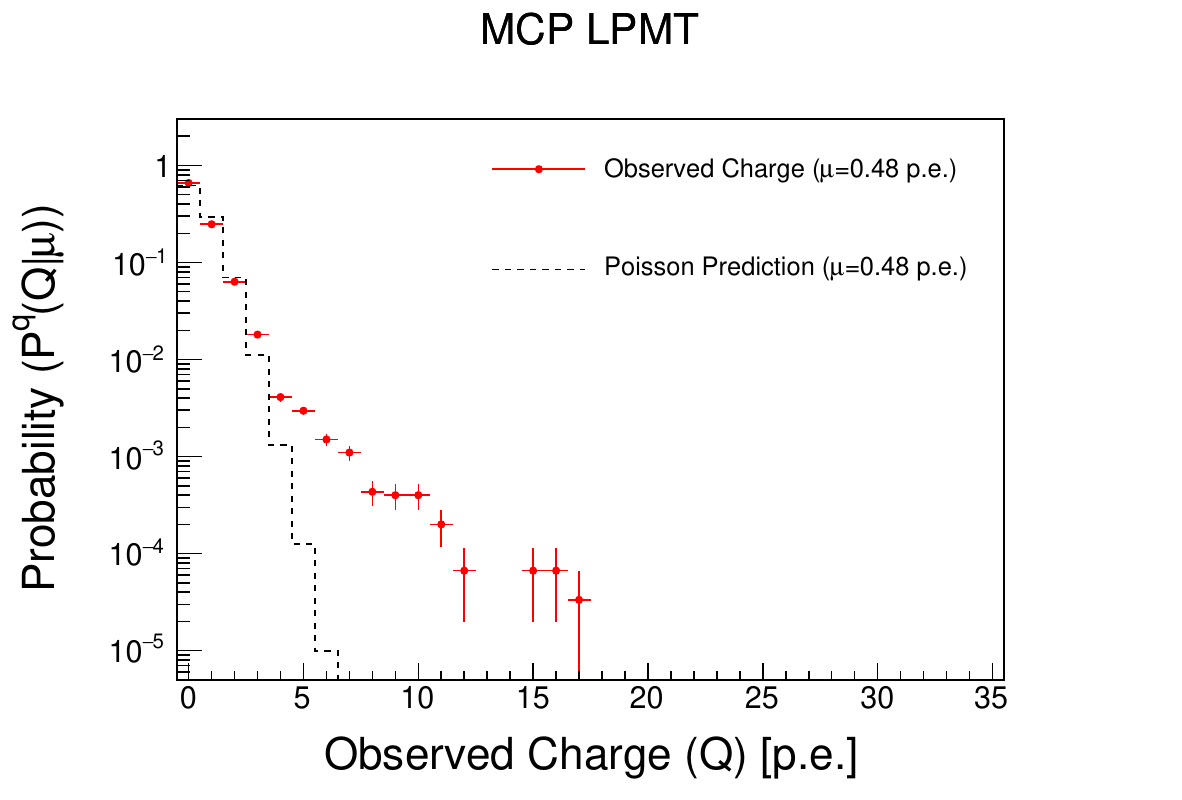}
\end{minipage} \\
\begin{minipage}{0.49\linewidth}
\centering
\includegraphics[width=1\linewidth]{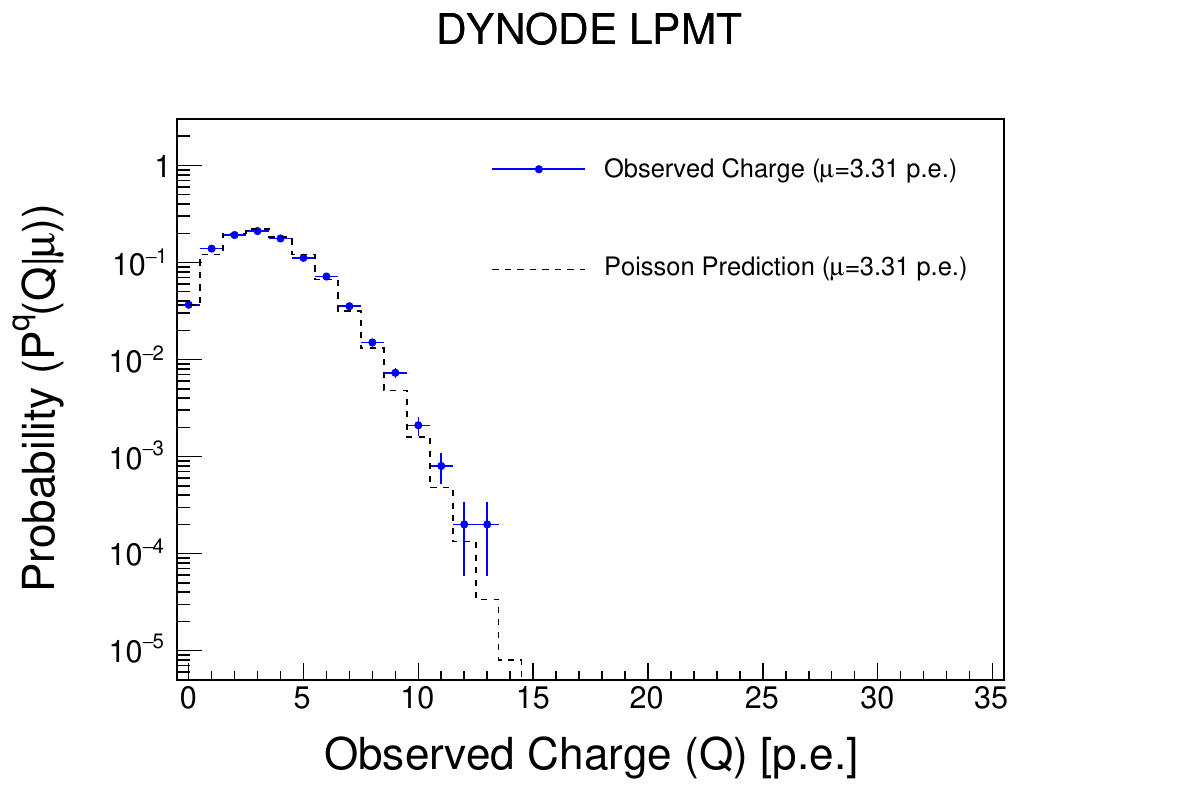}
\end{minipage}
\begin{minipage}{0.49\linewidth}
\centering
\includegraphics[width=1\linewidth]{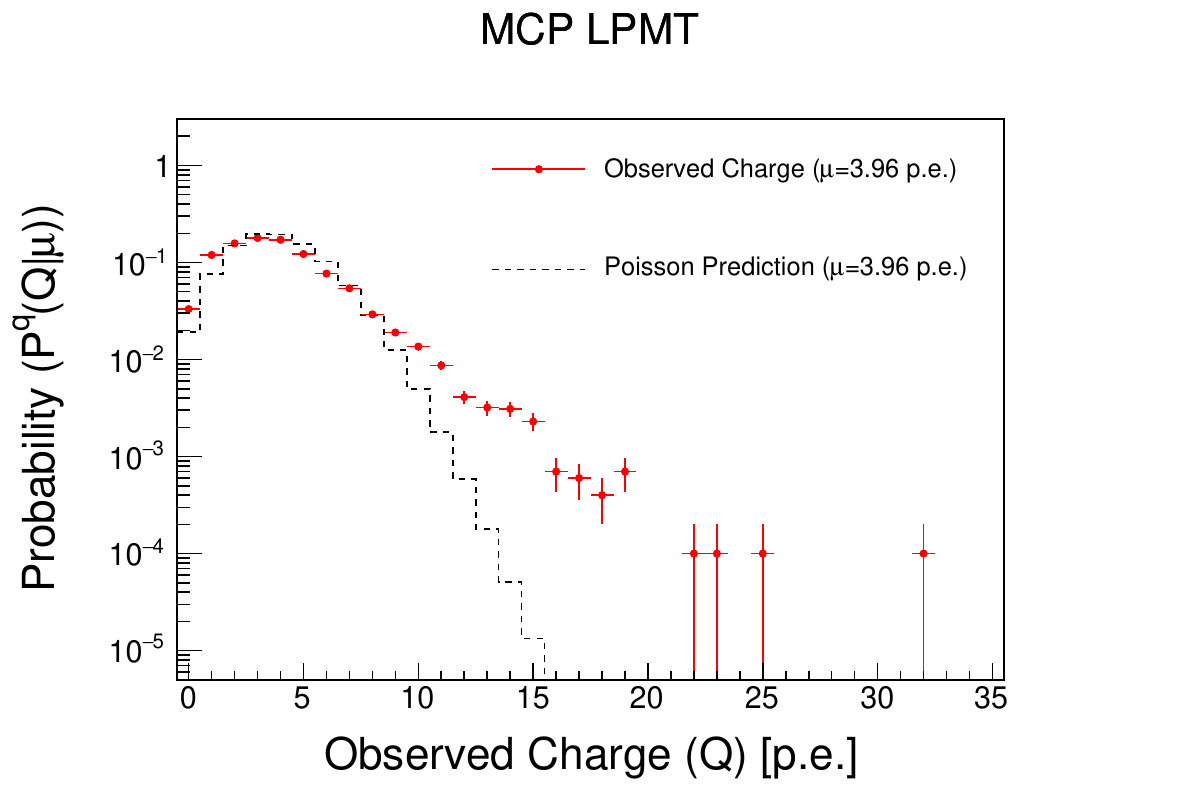}
\end{minipage} \\
\begin{minipage}{0.49\linewidth}
\centering
\includegraphics[width=1\linewidth]{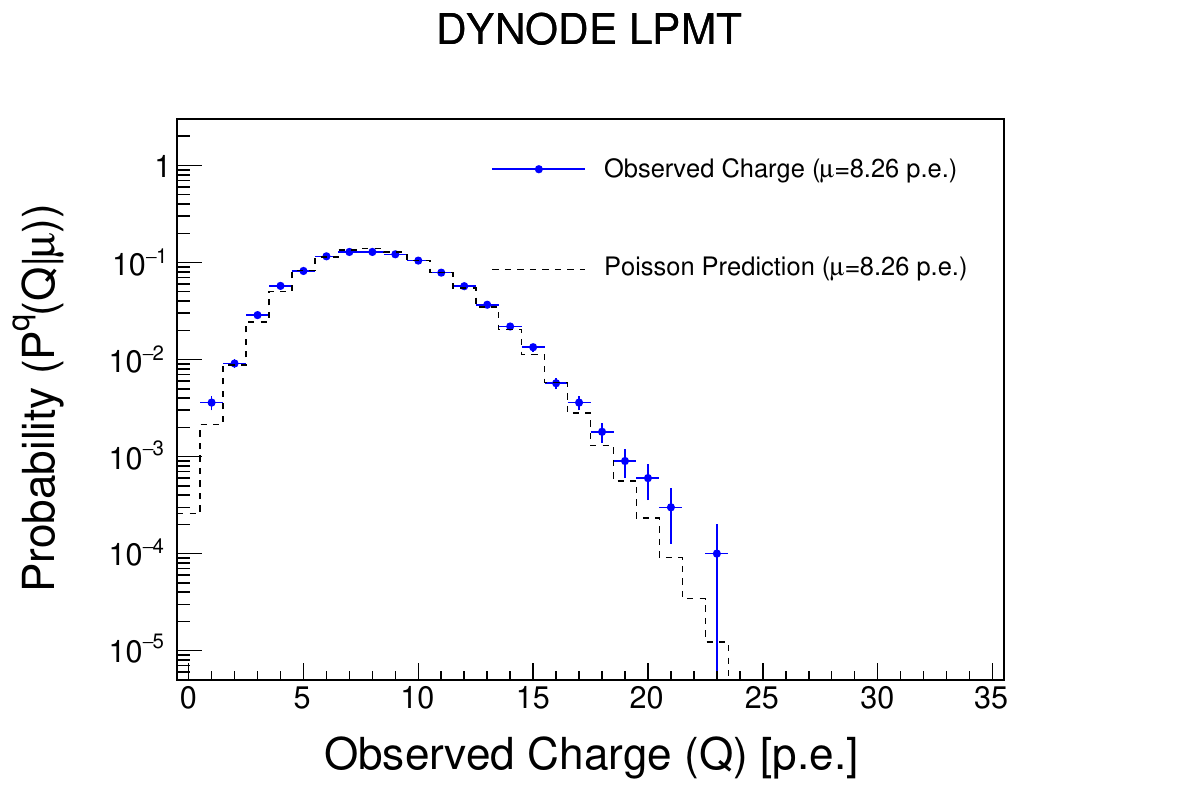}
\end{minipage}
\begin{minipage}{0.49\linewidth}
\centering
\includegraphics[width=1\linewidth]{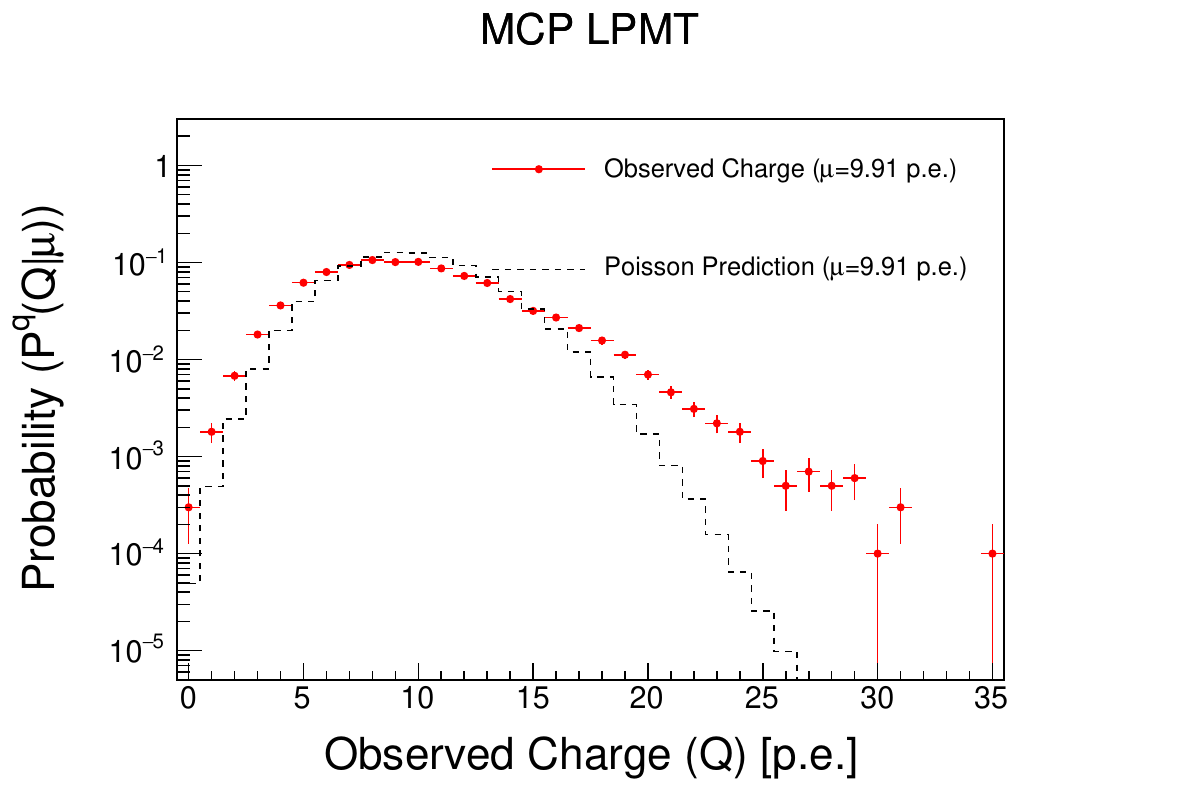}
\end{minipage}
\caption{
The left (right) plots show the normalized observed charge distributions for a Dynode (MCP) LPMT in different laser light intensity samples.
They are normalized by the number of laser events at each intensity.
The points correspond to the observed probabilities, and the dashed lines represent the Poisson distributions with the same mean value $\mu$.
Note that the value in the first bin represents the probability of not observing photoelectrons $(Q=0)$, while the second bin corresponds to the probability of observing $Q$ from 0 to 1.5~p.e ($0<Q<1.5$~p.e.).
The distribution for the MCP LPMT has a significant long-tail structure as reported in Ref.~\cite{Zhang:2021ruy}, which is strongly deviated from the pure Poisson prediction.
}
\label{fig:chargepdf1}
\end{figure*}

In photon counting experiments, the observed number of photons follows Poisson statistics, and the probability of observing $k$~photons with $\mu$ mean observed number of photons can be evaluated as follows:
\begin{align}
P(k|\mu) = \frac{\mu^{k}\exp\left(-\mu\right)}{k!}. \label{eq:chargepdf1}
\end{align}
However, in real detectors equipped with photosensors and electronics systems like JUNO, the observable quantity is not the number of photons, but the charge after the PMT amplification and subsequent electronics processes.
Therefore, the observed charge distribution (charge PDF, $(P^{q}(Q|\mu))$) does not follow a simple Poisson distribution (see Figure~\ref{fig:chargepdf1}), but a convolution of the Poisson and PMT single photoelectron charge distributions.
As shown in the right plots of Figure~\ref{fig:chargepdf1} and reported in Ref.~\cite{Zhang:2021ruy}, in the JUNO detector case, the MCP LPMTs are known to have a tail in their charge distribution, where the shape of the observed charge distribution deviates more prominently from the Poisson prediction than those for Dynode LPMTs.
This study employed the laser calibration samples to directly calibrate the charge distribution $(P^{q}(Q|\mu))$ with a given light luminosity $\mu$.

\subsubsection{Charge PDF construction in the low-charge region}
\label{subsubsec:chargepdf2}
\begin{figure*}
\begin{minipage}{0.49\linewidth}
\centering
\includegraphics[width=1\linewidth]{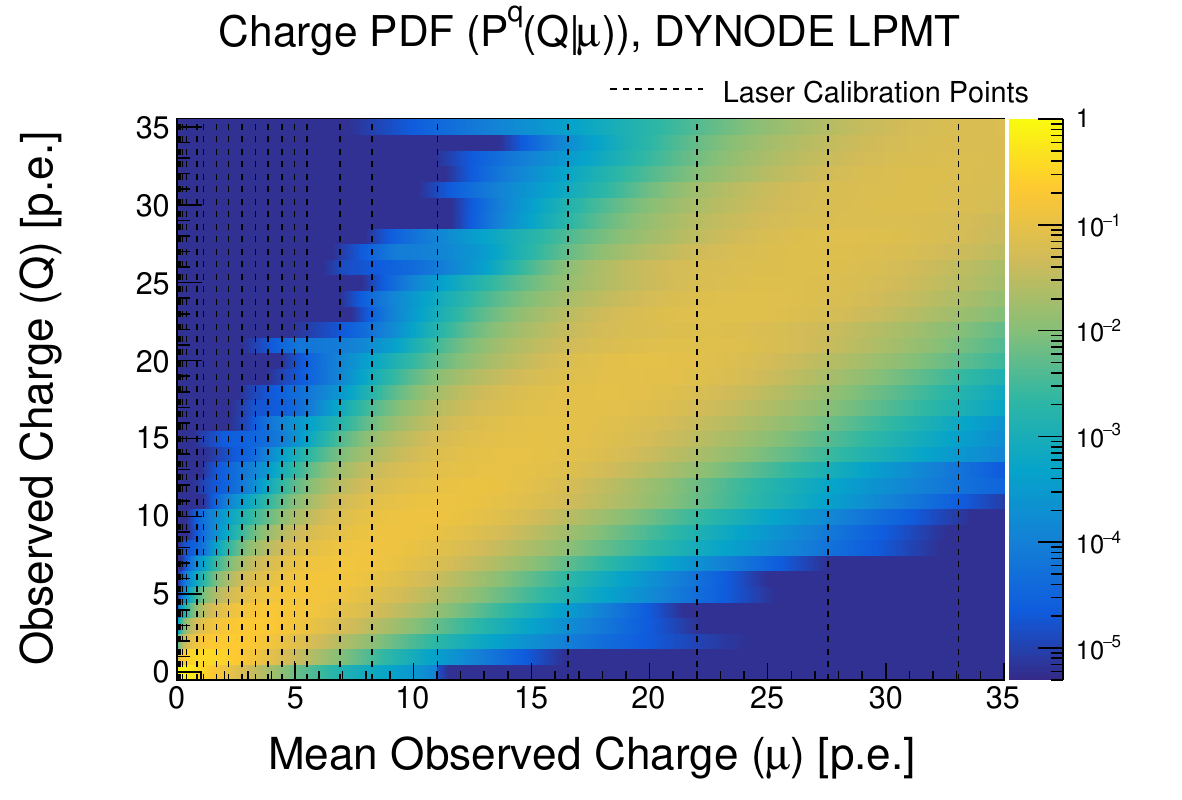}
\end{minipage}
\begin{minipage}{0.49\linewidth}
\centering
\includegraphics[width=1\linewidth]{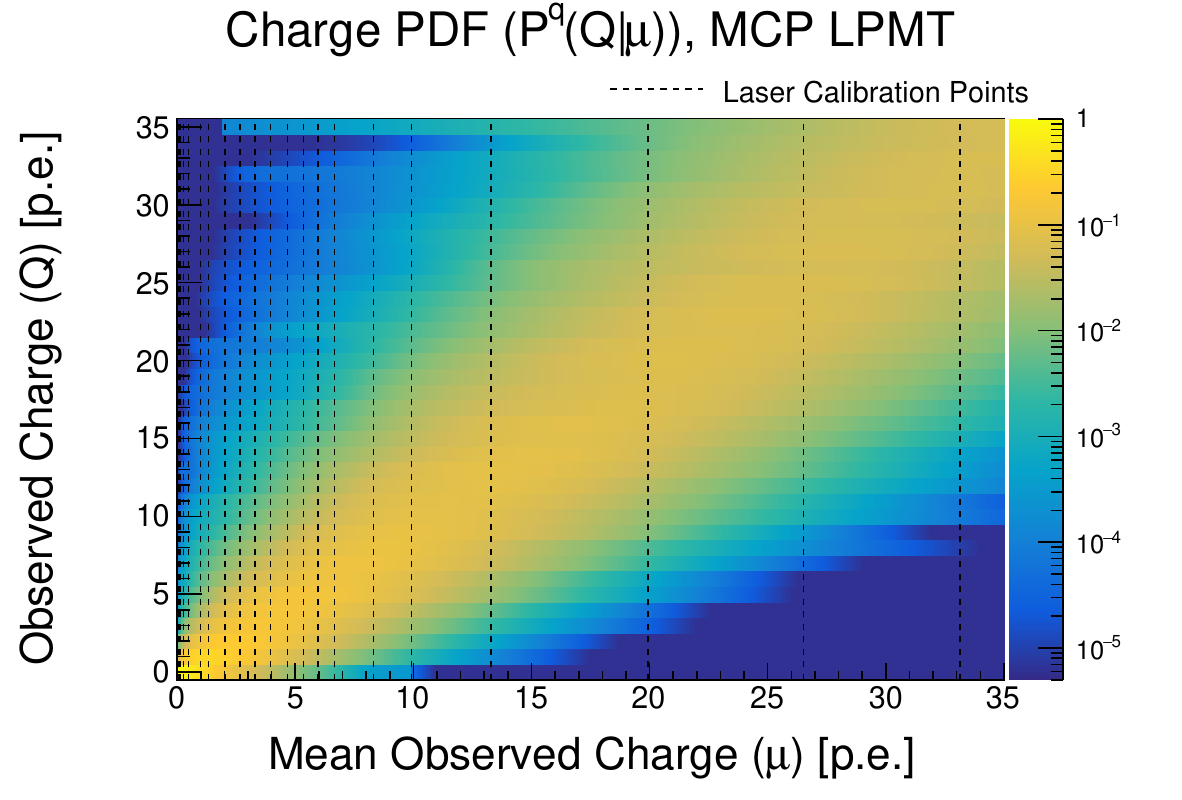}
\end{minipage}
\caption{
The left (right) plot shows the constructed charge PDF values for a Dynode (MCP) PMT, reflecting the probability of observing $Q$ with a given expected charge $\mu$.
Note that the value in the first bin in the vertical axis represents the probability of not observing photoelectrons $(Q=0)$, while the second bin corresponds to the probability of observing $Q$ from 0 to 1.5~p.e.
The vertical dashed lines represent the mean light intensities in various laser samples (calculated in Equation~(\ref{eq:chargepdf3})).
Charge PDF values between neighboring laser calibration points are interpolated by fitting the probabilities in each observed charge $(Q)$ slice with multiple polynomial functions of the mean observed charge ($\mu$).
Since the expected charge $\mu$ for most of the LPMTs in the reactor neutrino energy range is lower than 5~p.e. as shown in Figure~\ref{fig:chargemap1}, more laser calibration samples were generated in the lower-charge region to precisely capture the charge PDF in this region.
}
\label{fig:chargepdf4}
\end{figure*}

To directly calibrate the evolution of the LPMT charge distribution as a function of the average light intensity, a set of laser calibration samples was produced to cover low- to high-light intensity samples.
For every laser sample, the observed charge distribution was prepared, with the mean light luminosity (observed charge) at each LPMT evaluated as follows:
\begin{align}
\mu &= \frac{1}{N_{{\rm event}}}\sum_{k}^{N_{{\rm event}}}\left(Q_{k} -\mu_{{\rm dark}}\right), \label{eq:chargepdf3}
\end{align}
where $\mu$ is the mean observed charge at each LPMT in one laser sample, $N_{{\rm event}}$ is the number of events in the laser sample, $Q_{k}$ is the observed charge at each LPMT in the $k$th event in the laser sample, and $\mu_{{\rm dark}}$ is the dark noise contribution at this LPMT estimated in the same way as in Equation~(\ref{eq:chargemap4}). 
As shown in example plots in Figure~\ref{fig:chargepdf1}, representing the evolution of the observed charge distributions (charge PDFs) over different mean observed charges $(\mu)$ at each LPMT, charge PDF values at the observed charge $Q$ and a given $\mu$ value were directly obtained from individual laser samples, including the unhit probability ($Q=0$).
Continuous charge PDF tables were made by interpolating them using multiple polynomial functions of $\mu$ as illustrated in Figure~\ref{fig:chargepdf4}.
The fitting was performed on the negative logarithm ($-\log\left(P^{q}(Q|\mu)\right)$).
The bin at $Q=0$ was interpolated using a cubic spline function, while in the other $Q$ slices, the following function was used to fit $-\log\left(P^{q}(Q|\mu)\right)$ as a function of $\mu$:
\begin{eqnarray}
f(\mu) = -p_{\rm log}\log\mu+p_{0}+p_{1}\times\mu+p_{2}\times\mu^{2}.
\end{eqnarray}
Coefficients $(p_{\rm log}, p_{0}, p_{1}, p_{2})$ in the polynomial fit functions at each $Q$ slice are stored for each LPMT and are then looked up at the event reconstruction stage. \par

The observed charge ranges from 1~p.e. to over 100~p.e. for the reactor neutrino event energy in the JUNO detector.
However, it is computationally difficult to extend this charge PDF modeling up to the 100~p.e. level, due to the significant growth of the memory consumed for storing the polynomial coefficients.
Therefore, the charge PDF model introduced above was prepared up to 35~p.e. of the observed charge. 
Another charge PDF construction method was developed for the higher charge region, which is described in the following section.


\subsubsection{Charge PDF construction in the high-charge region}
\label{subsubsec:chargepdf3}
\begin{figure*}
\begin{minipage}{0.49\linewidth}
\centering
\includegraphics[width=1\linewidth]{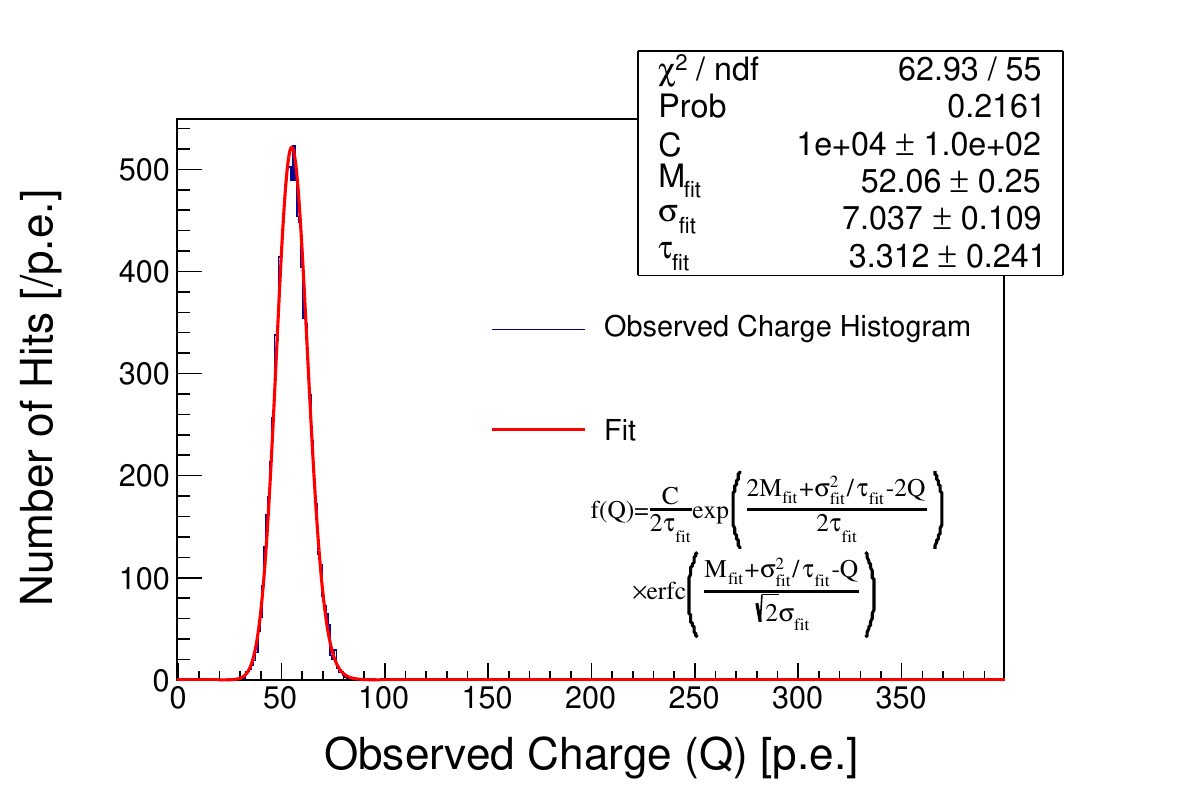}
\end{minipage}
\begin{minipage}{0.49\linewidth}
\centering
\includegraphics[width=1\linewidth]{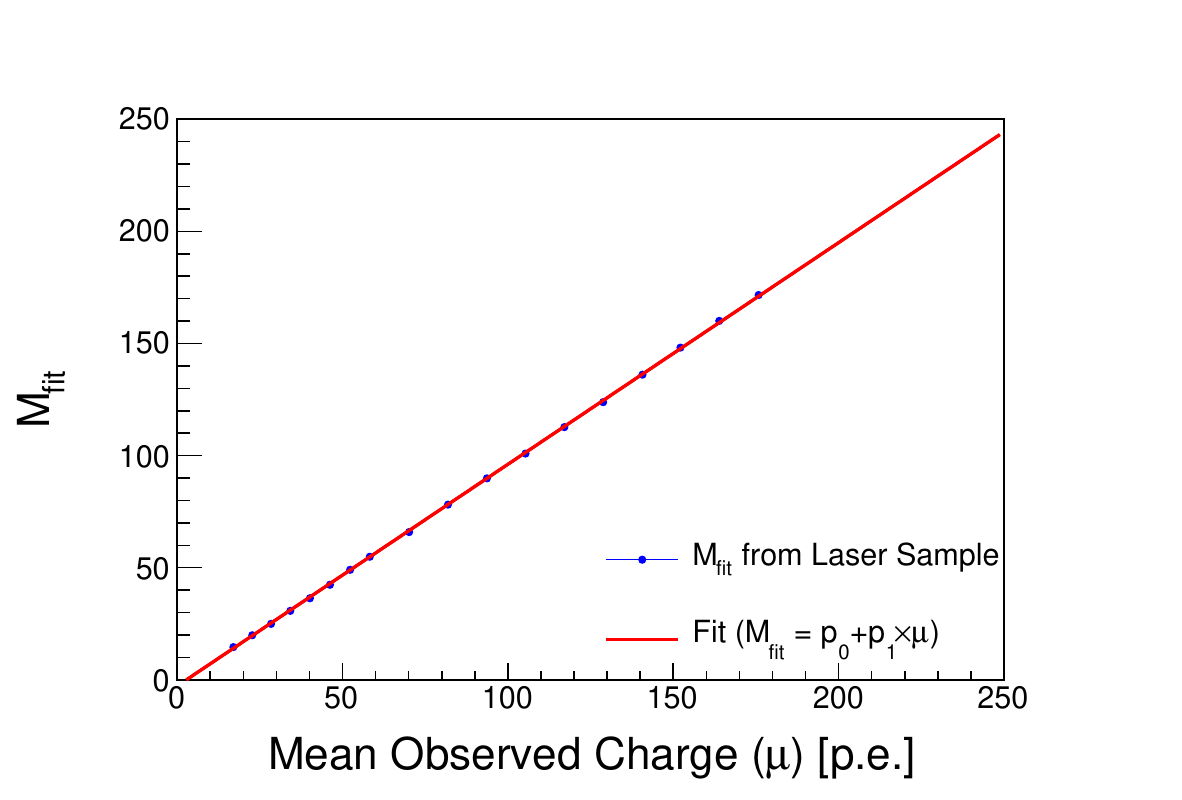}
\end{minipage} \\
\begin{minipage}{0.49\linewidth}
\centering
\includegraphics[width=1\linewidth]{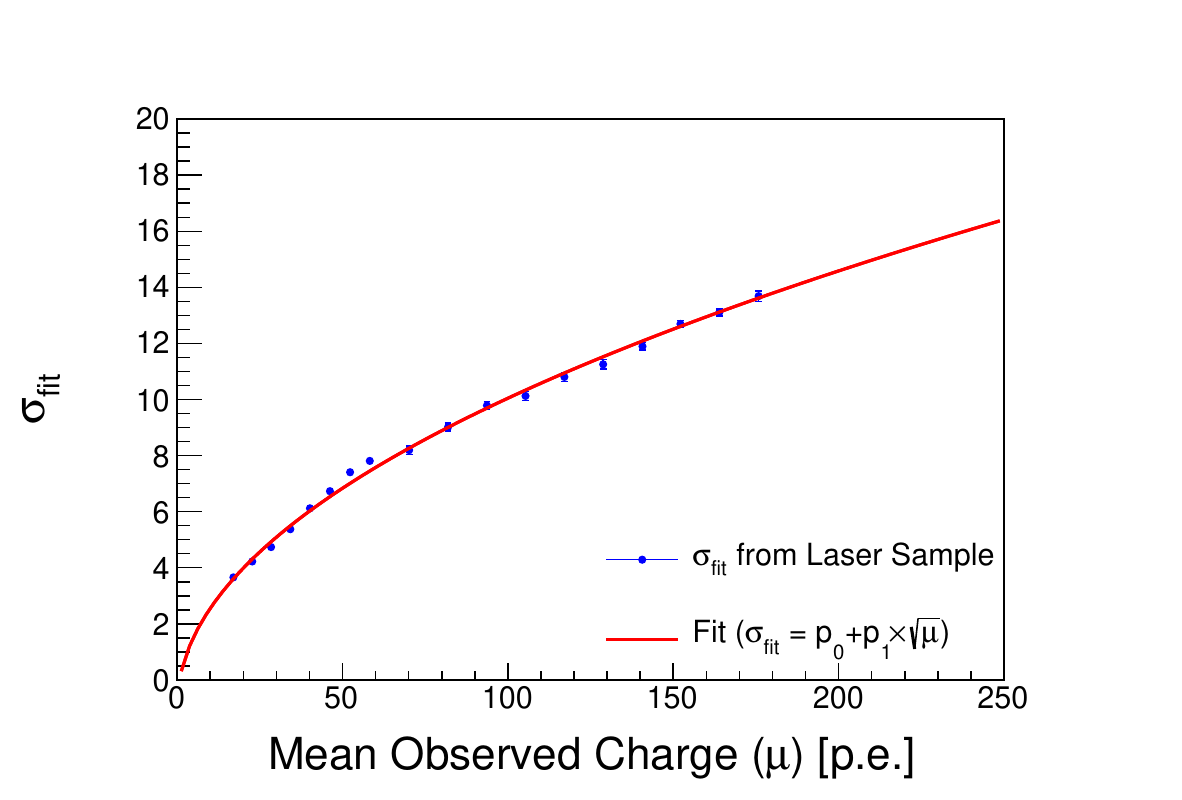}
\end{minipage}
\begin{minipage}{0.49\linewidth}
\centering
\includegraphics[width=1\linewidth]{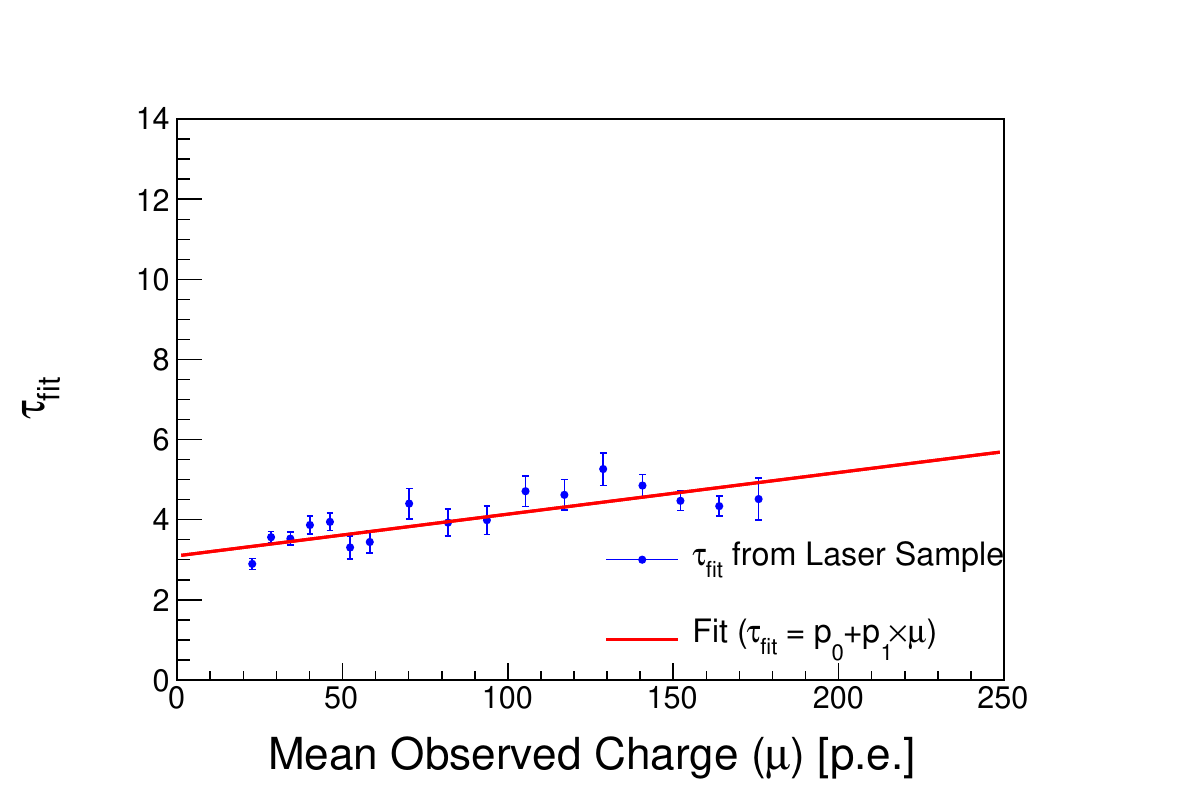}
\end{minipage}
\caption{
The top left plot shows the observed charge distribution at one Dynode LPMT with one of the high-intensity laser samples.
The solid red curve represents the exponentially modified Gaussian function used in the fit.
Evolution of the three parameters in the exponentially modified Gaussian function (see legend in the top left plot), $M_{\rm fit}$ (top right), $\sigma_{\rm fit}$ (bottom left), and $\tau_{\rm fit}$ (bottom right), is fitted as a function of mean observed charge $(\mu)$ as shown in red solid lines.
Vertical error bars in the plots for the three parameters represent the statistical uncertainty derived from the fitting shown in the top left plot.
}
\label{fig:chargepdf3}
\end{figure*}

As the number of detected photoelectrons increases, the charge distribution approaches a Gaussian distribution, as described by the central limit theorem.
The charge PDF tables for the LPMTs observing more than 35~p.e. were constructed utilizing this feature.
First, the observed charge distribution at each LPMT in each high-intensity laser sample was fitted with an exponentially modified Gaussian function (see Figure~\ref{fig:chargepdf3}).
This fitting function, after being normalized, represents the probability density function, reflecting the probability of observing a charge of $Q$ with a given mean observed charge of $\mu$, which is obtained from Equation~(\ref{eq:chargepdf3}).
After the fitting was applied to the charge distribution for individual LPMTs in each laser sample, the evolution of the three fitting parameters ($M_{\rm fit}, \sigma_{\rm fit}, \tau_{\rm fit}$ in the top left plot in Figure~\ref{fig:chargepdf3}) as a function of mean observed charge $(\mu)$ was formulated using analytical functions as shown in Figure~\ref{fig:chargepdf3}.
Coefficients ($p_{0}, p_{1}$ in Figure~\ref{fig:chargepdf3}) describing the relationship between each fitting parameter and $\mu$ were saved for each LPMT and are then looked up at the event reconstruction stage.

\section{Reconstruction performance}
\label{sec:performance}
The reconstruction algorithms with the prepared input tables based on the calibration samples were applied to independent simulation samples to evaluate their reconstruction performance.
As described in Table~\ref{tab:simulation2}, positrons, $\alpha$-particles, and fast-neutrons of various energies were generated for this test.
The performance of the vertex and energy reconstruction algorithms was evaluated with the positron samples in Sections~\ref{subsec:performance1} and~\ref{subsec:performance2} since the algorithm presented in this paper will be applied to the reactor neutrino events in JUNO.
The $\alpha$-particle and fast-neutron samples were additionally analyzed to evaluate the particle identification performance discussed in Section~\ref{subsec:performance3}.

\subsection{Vertex reconstruction performance}
\label{subsec:performance1}
\begin{figure*}
\begin{minipage}{0.49\linewidth}
\centering
\includegraphics[width=1\linewidth]{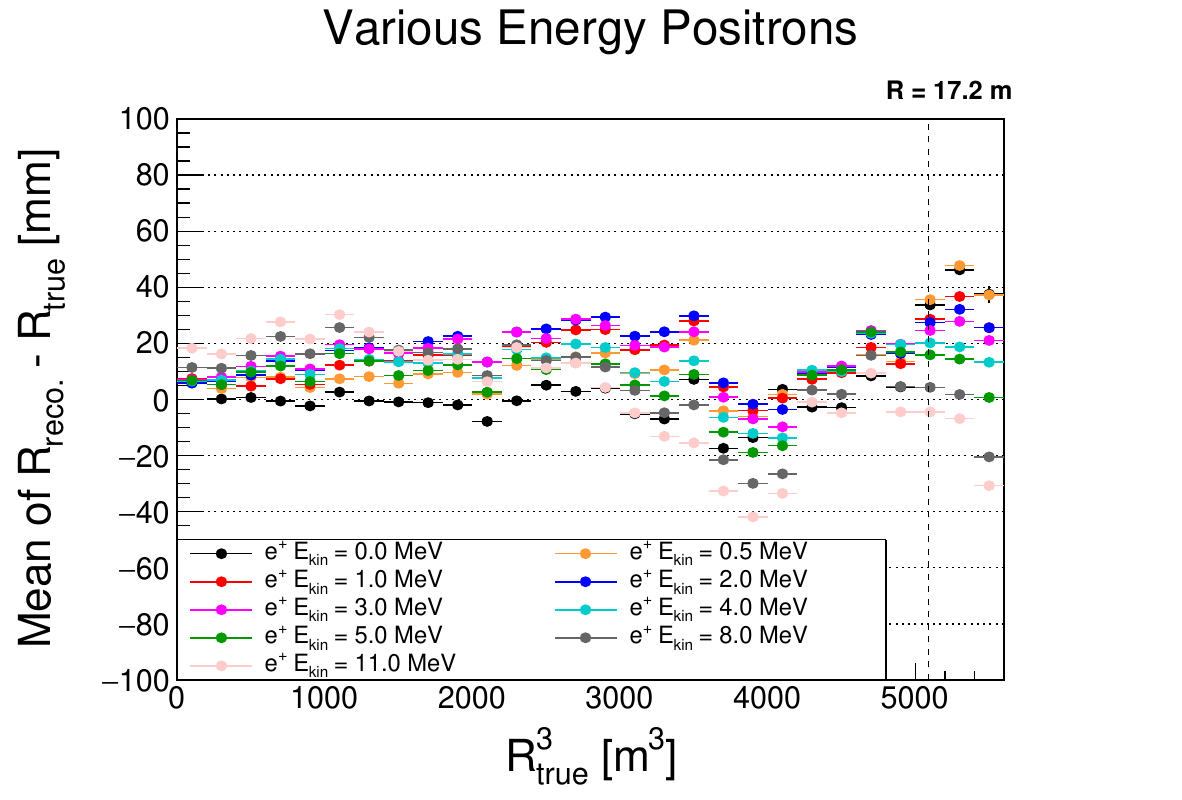}
\end{minipage}
\begin{minipage}{0.49\linewidth}
\centering
\includegraphics[width=1\linewidth]{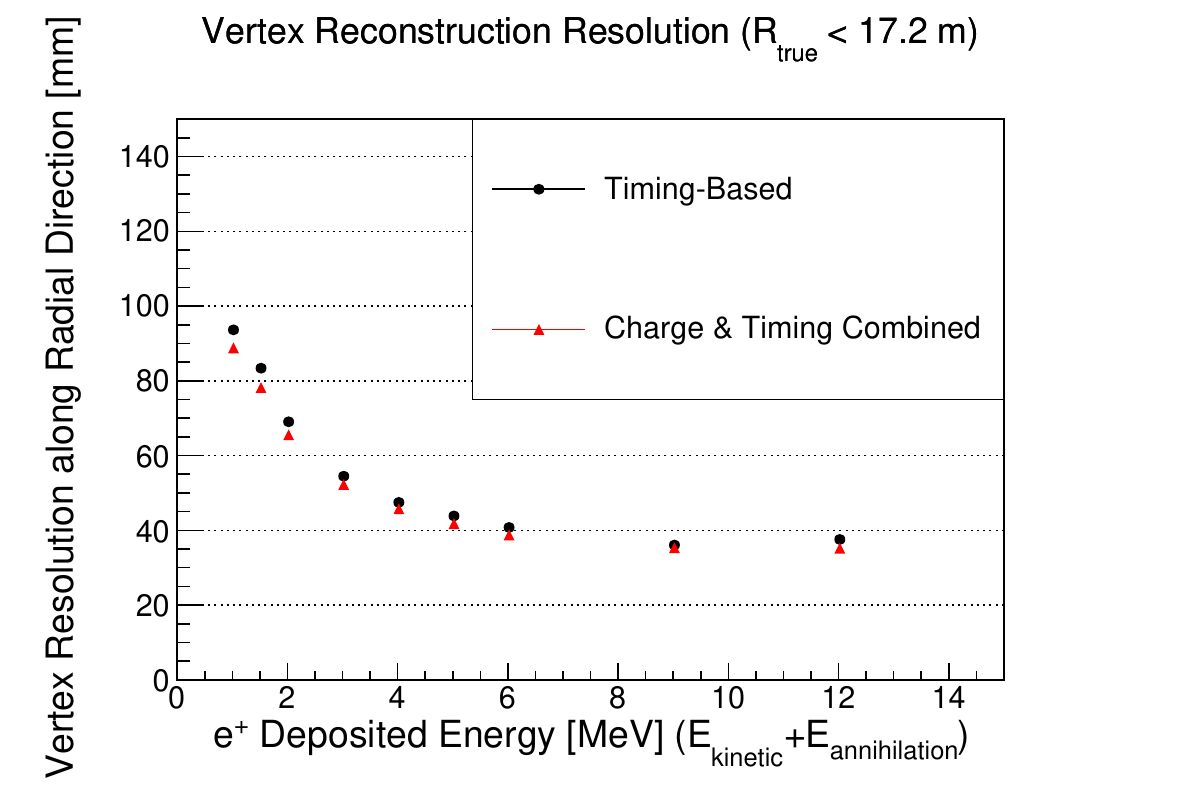}
\end{minipage}
\caption{
The left plot shows the mean vertex reconstruction bias dependence as a function of the cubic of the true event radial position $(R^{3})$ for uniformly distributed positrons of various energies using the charge and timing combined algorithm.
In the vertical axis title, $R_{\rm reco.}$ $(R_{\rm true})$ denotes the reconstructed (true) vertex radial position.
The vertical dashed line at $R^{3}=17.2^{3}\approx5088~{\rm m}^{3}$ indicates the fiducial volume boundary of the JUNO detector.
The right plot shows the vertex reconstruction resolutions against the positron deposited energy, and $E_{\rm annhilation}$ denotes the positron annihilation energy, 1.022~MeV.
The resolution here was obtained by a Gaussian fitting $(\sigma)$ to the $(R_{\rm reco.}-R_{\rm true})$ distribution for each position sample.
The black filled circle (red triangle) in the right plot shows the performance of the timing-only (charge and timing combined) algorithm.
}
\label{fig:performance1}
\end{figure*}

The vertex reconstruction performance was evaluated by comparing the reconstructed vertex position to the average position of the particle energy deposition (simulation truth information).
This comparison was made on an event-by-event basis, and the mean difference between the reconstructed event radial position $(R_{\rm reco.})$ and true event radial position $(R_{\rm true})$ was estimated because the bias along the radial direction has more influence on the energy reconstruction process afterwards.
The left plot in Figure~\ref{fig:performance1} shows the radial reconstruction bias as a function of the detector radial region for the positron samples with different kinetic energies using the charge and timing combined algorithm (Equation~(\ref{eq:eventreco4})).
It confirms that the vertex bias along the detector radial direction is smaller than 4~cm throughout the liquid scintillator volume.
The vertex reconstruction resolution along the detector radial direction was evaluated by applying a Gaussian fit to the $(R_{\rm reco.}-R_{\rm true})$ distribution made from events within $R=17.2$~m, the JUNO fiducial volume boundary.
The right plot of Figure~\ref{fig:performance1} shows the vertex resolutions against the positron deposited energy, where the charge and timing combined algorithm shows better vertex resolutions in the whole energy range, compared to the timing-only algorithm, since more information is exploited in the combined algorithm.
The vertex resolution for positrons with a kinetic energy of 0~MeV (two 511~keV $\gamma$-rays from the ${\rm e}^{\pm}$ annihilation) is estimated to be around 9~cm using the charge and timing combined algorithm.
As the positron energy increases, more photons are observed in the detector, leading to the vertex resolution improvement up to around 10~MeV.
At even higher energies, the vertex reconstruction resolution does not improve as the positron energy increases.
It is confirmed that this is due to the fact that the residual time distribution from higher-energy positrons has more discrepancies from the input timing PDF produced by the $\gamma$-ray samples.
This is considered to arise from the difference that higher-energy positrons produce more localized energy deposits on the millimeter scale, whereas $\gamma$-rays scatter electrons along their paths over distances on the order of 10~centimeters.


\subsection{Energy reconstruction performance}
\label{subsec:performance2}
The light yield in liquid scintillator detectors is generically not proportional to the particle's deposited energy due to the quenching effect~\cite{Birks:1964zz} and Cherenkov photon contribution~\cite{Frank:1937fk}.
This non-linear dependence of the light yield on the particle energy makes it difficult to compare the reconstructed energy to the true deposited energy. 
Thus, this paper focuses on the uniformity of the reconstructed energy inside the detector and energy resolution.
The calibration strategy for the non-linear detector response was discussed in Refs.~\cite{JUNO:2020xtj, Yu:2022god}, and the methods discussed therein to calibrate the non-linearity of the liquid scintillator response can be applied after this energy reconstruction. \par
\begin{figure*}
\begin{minipage}{0.49\linewidth}
\centering
\includegraphics[width=1\linewidth]{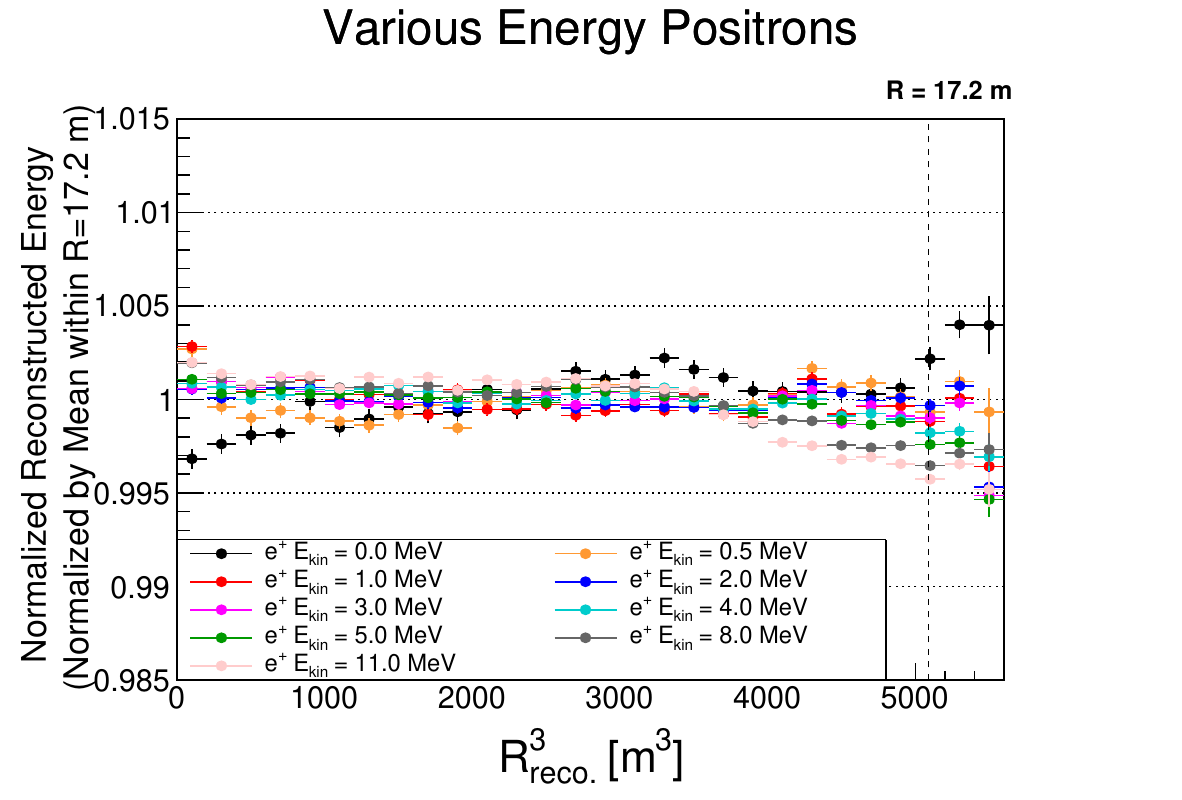}
\end{minipage}
\begin{minipage}{0.49\linewidth}
\centering
\includegraphics[width=1\linewidth]{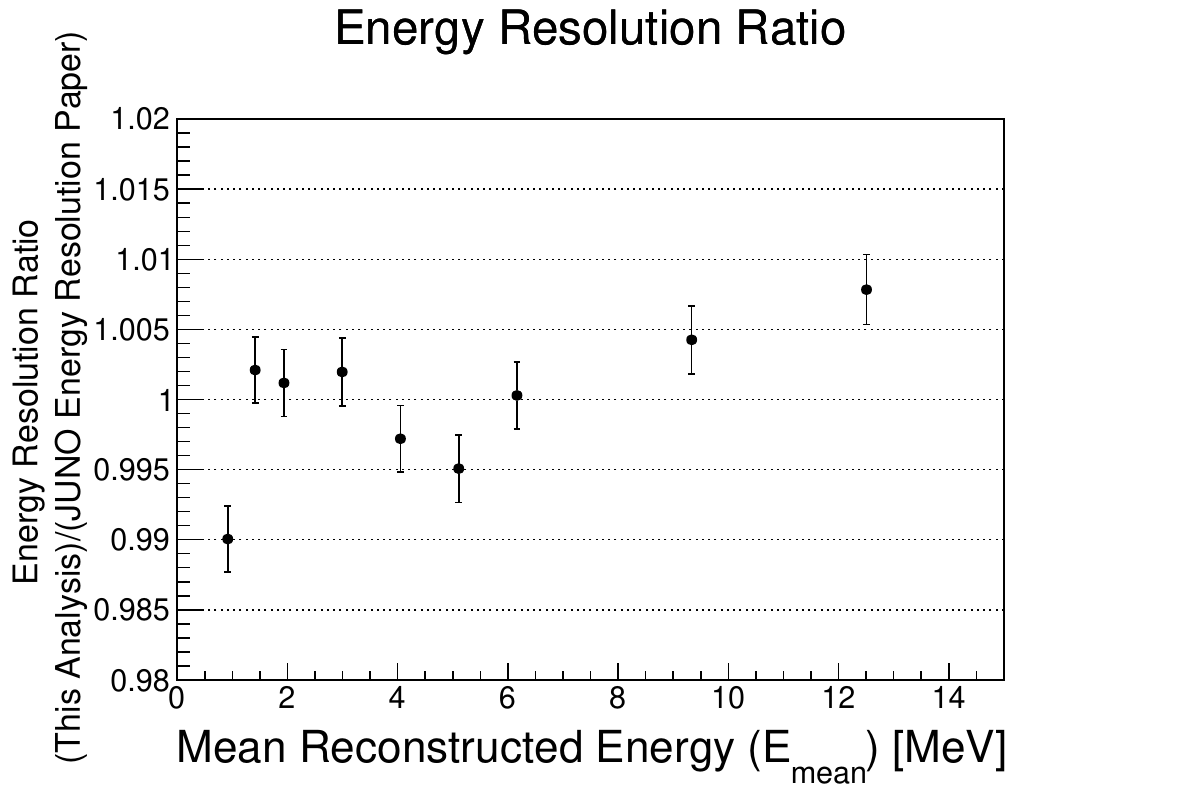}
\end{minipage}
\caption{
The left plot shows the normalized reconstructed energy distribution as a function of the cubic of the reconstructed event radial position $(R_{\rm reco.}^{3})$ for uniformly distributed positron samples of various energies.
The values in the vertical axis are normalized by the average reconstructed energy within $R=17.2$~m.
The vertical dashed line at $R^{3}=17.2^{3}\approx5088~{\rm m}^{3}$ indicates the fiducial volume boundary of the JUNO detector.
The right plot shows the evaluated energy resolutions using the present algorithm relative to the JUNO official ones presented in Ref.~\cite{JUNO:2024fdc}.
The resolution was obtained by a Gaussian fit ($\sigma$/$E_{\rm mean}$) of the reconstructed energy distribution made from the uniformly distributed positron events within $R=17.2$~m.
Vertical error bars represent the statistical uncertainty of the fitting.
}
\label{fig:performance2}
\end{figure*}

The energy reconstruction algorithm presented in Section~\ref{subsec:eventreco2} was applied to the uniformly distributed positron events of various energies.
Particles produced near the detector edges can escape outside of the liquid scintillator with non-zero momentum, distorting the true deposited energy as well as reconstructed energy distribution.
This energy leakage needs to be carefully and independently considered in the physics analysis.
This study focuses on the full energy-deposition events to decouple the energy reconstruction quality from this issue.
The left plot of Figure~\ref{fig:performance2} shows the normalized reconstructed energy as a function of event position. 
The value in the vertical axis was obtained from the peak position of the Gaussian fitting to the reconstructed energy distribution in each detector region.
In addition, these values are normalized by the mean reconstructed energy obtained from the events within $R=17.2$~m, for each mono-energetic positron sample. 
The non-uniformity of the reconstructed energy is confirmed to remain within a 0.5\% level inside the detector, comparable with the result based on the JUNO standard algorithm~\cite{JUNO:2024fdc}.
The energy resolution for positron events was estimated by fitting a Gaussian to the reconstructed energy distribution made by the positron events within $R=17.2$~m. 
The right plot of Figure~\ref{fig:performance2} shows the evaluated energy resolutions relative to the JUNO official ones presented in Ref.~\cite{JUNO:2024fdc} estimated with another event reconstruction algorithm detailed in~Refs.~\cite{Huang:2021baf, Huang:2022zum} based on $^{68}$Ge calibration samples across the detector.
The main difference between the two algorithms lies in the methodology for constructing the charge map. 
The algorithm presented in this paper utilizes $\gamma$-rays originating from neutron captures that are randomly produced within the detector, while the algorithm described in the official JUNO publication relies on calibration samples obtained from more than 290 fixed points using various calibration source deployment systems.
Despite being developed using substantially fewer fixed-point calibration samples, the present algorithm demonstrates comparable performance, with the energy resolution differing by less than 1\% relative to the official results.

\subsection{Particle identification performance}
\label{subsec:performance3}
\begin{figure}
\includegraphics[width=0.48\textwidth]{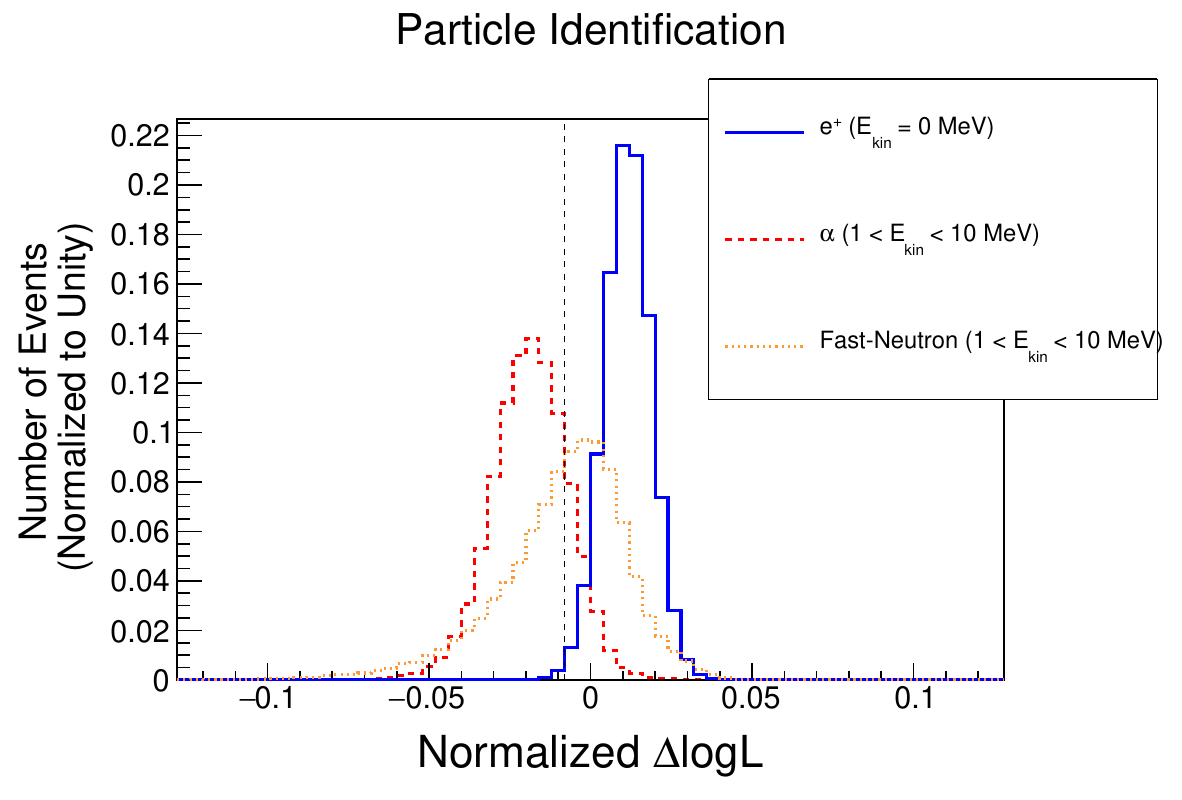}
\caption{
The normalized delta log-likelihood distribution for the positron with $E_{\rm kin}=0$~MeV (blue solid), $\alpha$-particle (red dashed), and fast-neutron (orange dotted) samples.
The positive value in these distributions indicates $\gamma$-ray-like events while the negative value indicates fast-neutron-like events.
The vertical dashed line at -0.008 corresponds to the threshold value for particle identification as a benchmark.
}
\label{fig:performance3}
\end{figure}

In the reactor neutrino analysis in JUNO, positron events are regarded as signal while $\alpha$-particles and fast-neutrons are considered to be background, and this algorithm aims to separate them.
After vertex reconstruction, the negative log-likelihood value for the fast-neutron assumption was evaluated and compared to the minimized negative log-likelihood value based on the $\gamma$-ray assumption in the vertex reconstruction process.
The following normalized delta log-likelihood is defined:
\begin{align}
\left(\Delta\log L\right)_{\rm norm.} &= \frac{\log L(\text{Fast-neutron})-\log L(\text{$\gamma$-ray})}{N_{\rm fired~LPMTs}},
\end{align}
where $N_{\rm fired~LPMTs}$ is the number of fired LPMTs in each event and acts as the normalization factor to mitigate the energy dependence of $\Delta\log L$.
For fast-neutron and $\alpha$-particle events, by definition, this quantity tends to be negative, while it approaches positive values for electron, positron, and $\gamma$-ray events.
As the light emission time profile from higher-energy positrons is sharper and more distinct compared to that of $\alpha$-particles and fast-neutrons, the lowest-energy positron sample is the most difficult sample to separate from $\alpha$-particles and fast-neutrons.
Therefore, this paper focuses on the separation performance between stationary positrons ($E_{\rm kin}=0$~MeV) and $\alpha$-particles/fast-neutrons.
Figure~\ref{fig:performance3} shows the delta log-likelihood distributions for the uniformly distributed positrons with $E_{\rm kin} = 0$~MeV, $\alpha$-particles with $1 < E_{\rm kin} < 10$~MeV, and fast-neutrons with $1 < E_{\rm kin} < 10$~MeV, confirming that this algorithm has a certain particle identification power.
The threshold value to separate the $\gamma$-ray-like events and fast-neutron-like events can be tuned in higher-level analyses since desired signal efficiencies may vary depending on the physics objective.
A benchmark threshold was set at -0.008 to evaluate the particle separation power in this study.
This threshold condition was determined to keep the positron signal misidentification rate less than 1\%, and roughly 80\% (45\%) of the $\alpha$ (fast-neutron) events are correctly categorized as fast-neutron-like background events.
The more efficient separation power for the $\alpha$ sample is achieved compared to the fast-neutron sample because $\alpha$-particles have higher $dE/dx$ than protons scattered by fast-neutrons, resulting in even slower scintillation photon emissions.

\section{Conclusion}
\label{sec:conclusion}
This paper presents details of a neutron source-based event reconstruction in large liquid scintillator detectors, including the event vertex position, event energy reconstruction, and particle identification algorithms, as well as the methods for preparing input tables based on the simulated $^{241}$Am$^{13}$C source calibration, uniformly distributed neutron and laser calibration samples. 
For the timing PDF tables, we found that the neutron and $\gamma$-ray travel length in the liquid scintillator had a substantial impact on the timing distribution and developed a method to iteratively mitigate this influence by referring to the reconstructed event position instead of relying on the source deployment position.
The improvement in the timing PDF tables significantly reduces the vertex reconstruction biases.
The vertex reconstruction bias is kept within 4~cm level throughout the detector and resolution for events with around 1~MeV energy deposition is estimated to be approximately 9~cm. \par
To utilize the charge and hit information at each PMT in the event reconstruction, uniformly distributed 2.2~MeV $\gamma$-ray events from neutron captures and laser calibration events were employed to calibrate detector response, such as the mean expected charge at each PMT with a given event vertex position and the LPMT charge response (charge PDF).
Energy reconstruction is done by comparing the expected charge and observed charge at each LPMT through the charge PDF tables, as well as computing the expected photon detection probability at each SPMT.
Reconstruction tests on positron samples of various energies confirm that the reconstructed energy uniformity is better than 0.5\% and the energy resolution is comparable to the JUNO publication (within 1\% level).
Finally, particle identification performance was estimated.
The particle identification is done by comparing the observed timing distribution to the two prepared timing PDF tables made from the $\gamma$-ray and fast-neutron samples from the $^{241}$Am$^{13}$C source calibration events.
As a result, 80\% (45\%) of the uniformly distributed $\alpha$ (fast-neuron) events are correctly identified as the fast-neutron-like events, while the misidentification rate for the positron remains less than 1\%. \par
This study was performed based on simulation samples in JUNO but represents a data-driven approach by maximally using the calibration and background samples.
With the imminent data-taking in JUNO, these methods can be tested and improved on the real data.
The application of these methods can be extended to other large liquid scintillator detectors as well.

\begin{acknowledgements}
This work is supported by the National Key Research and Development Program of China (Grant no. 2023YFA1606104), and by National Science Foundation of China for International Young Scientists (Grant Number 12250410235).
Y. M. and J. H. thank the sponsorship from the Yangyang Development Fund.
A. T. and I.M.B. also thank K. C. Wong Educational Foundation for their financial support.
The authors appreciate all JUNO collaborators for their valuable feedback, especially the members of the reconstruction working group, as well as those who have contributed to the development and testing of the JUNO software.
\end{acknowledgements}

\bibliographystyle{spphys}       
\bibliography{reference}   


\end{document}